\documentclass{article}%
\usepackage{amsmath}
\usepackage{graphicx}
\usepackage{amsfonts}
\usepackage{amssymb}%
\newtheorem{theorem}{Theorem}
\newtheorem{acknowledgement}[theorem]{Acknowledgement}

\begin{document}

\title{String theory, the crisis in particle physics\\and the ascent of metaphoric arguments \\{\small Dedicated to the memory of Juergen Ehlers}\\IJMPD \textbf{17}, (2008) 2373-2431}
\author{Bert Schroer\\CBPF, Rua Dr. Xavier Sigaud 150 \\22290-180 Rio de Janeiro, Brazil\\and Institut fuer Theoretische Physik der FU Berlin, Germany}
\date{September 2008}
\maketitle
\tableofcontents

\begin{abstract}
This essay presents a critical evaluation of the concepts of string theory and
its impact on particle physics. The point of departure is a historical review
of four decades of string theory within the broader context of six decades of
failed attempts at an autonomous S-matrix approach to particle theory.

The central message, contained in sections 5 and 6, is that string theory is
not what its name suggests, namely a theory of of objects in spacetime whose
localization is string- instead of point-like. Contrary to popular opinion the
oscillators corresponding to the Fourier models of a quantum mechanical string
do not cause a stringlike spatial extension of the object under discussion and
neither does the "range space" of a chiral conformal QFT acquire the
interpretation of string-like localized quantum matter.

Rather string theory represents a solution of a problem which enjoyed some
popularity in the 60s: find a principle which, similar to the SO(4,2) group in
the case of the hydrogen spectrum, determines an infinite component wave
function with a (realistic) mass/spin spectrum. Instead of the group theory
used in the old failed attempts, it creates this mass/spin spectrum by
combining an \textit{internal oscillator quantum mechanics} with a pointlike
localized quantum field theoretic object, i.e. the mass/spin tower "sits" over
one point and does not arise from a wiggling string in spacetime.

The widespread acceptance of a theory whose interpretation has been based on
metaphoric reasoning had a corroding influence on particle theory, a point
which will be illustrated in the last section with some remarks of a more
sociological nature. These remarks also lend additional support about
observations on connections between the discourse in particle physics and the
present Zeitgeist of the post cold war period made in the introduction.

\end{abstract}

\section{Symptoms of a crisis in the foundations of particle theory}

There can be no doubt that after almost a century of impressive success,
fundamental physics is in the midst of a deep crisis. Its epicenter is in
particle theory, but its repercussions may even influence the direction of
experimental particle physics and affect adjacent areas of fundamental
research which traditionally used innovative ideas of quantum field theory
(QFT). They also led to quite bizarre ideas about the philosophy underlying
fundamental sciences, which partially explains why they attracted considerable
attention beyond the community of specialists in particle physics.

One does not have to be a physicist in order to be amazed when reputable
members of the particle physics community \cite{Suss} recommend a paradigmatic
change away from the observation based setting of physics which, since the
time of Galileio, Newton, Einstein and the protagonists of quantum theory has
been the de-mystification of nature by mathematically formulated concepts with
experimentally verifiable consequences. The new message, which has been formed
under the strong influence of string theory, is that it is scientifically
acceptable to use ones own existence in reasonings about matters of
theoretical physics, even if this leads to a vast collection of in principle
unobservable concepts such as \textit{multiverses} and \textit{parallel
worlds}. This new physics excepts metaphors but calls them principles, as e.g.
the \textit{anthropic principle}; its underlying philosophy resembles a
religious faith with its unobservable regions of heaven and hell rather than
physics, as we know it since the times of Galileio. It certainly amounts to a
rupture with traditional natural sciences and the philosophy of enlightenment.
Despite assurances to the contrary, it looks like an avatar of
\textit{intelligent design}.

Instead of \textquotedblleft cogito ergo sum\textquotedblright\ of the
rationalists, the new \textit{anthropic} maxim coming from this new doctrine
attaches explanatory power to its inversion: "I exist and therefore things are
the way they are", since otherwise I would not exist. Its main purpose is to
uphold the uniqueness of the string theorists dream of a theory of everything
(TOE). Even with an enormous number of string solutions with different
fundamental laws and fundamental constants, the use of the anthropic principle
permits to uphold uniqueness by claiming that we are living in a multiverse
consisting of as many universes as it needs to account for all string
solutions; in this way one is able to claim that these describe actually
existing but inaccessible and invisible parallel worlds or multiverses of a
unique TOE. In this context anthropic reasoning is not meant as a temporary
auxiliary selective device, pending better understanding and further
refinements of the theory, but rather as a way to uphold the TOE status of
string theory. To physicists outside the string community, the logic behind
this doctrine resembles the "if you cannot solve a problem then enlarge it"
motto of some politicians.

To demonstrate the physical relevance of string theory in this anthropic
setting it would suffice to show that there is one solution which looks like
our universe; but whereas the number of solutions has been estimated, nobody
has an idea how to arrange such a search. How can one find something in a
haystack if one does not even know how to characterize our universe in terms
of moduli and other string-theoretic notions which distinguish a reference
state (the string theory vacuum)?

To be fair, the anthropic dogma of a \textit{multiverse} instead of a universe
i.e. the belief that all these different solutions with quantum matter obeying
different laws (including different values of fundamental constants) exist and
form "the landscape" \cite{Suss}, is not shared by all string theorists. \ 

Such a picture is still confined to a vocal and influential minority in
particle theory, but it is not difficult to notice a general trend of moving
away from the traditional scientific setting based on autonomous physical
principles, towards more free-wheeling metaphoric consistency arguments.
Ironically some of the new aggressive science-based atheists are strong
defenders of the metaphors about the string-inspired multiverses.

The ascent of this metaphoric approach is fueled by the increasing popularity
of string theory and the marketing skill of its proponents to secure its
funding. This, and certainly not the extremely meager physical results, is
what at least partially accounts for its present dominant status in particle
theory. Whereas the attraction it exerts on newcomers in physics is often
related to career-building, the attention it receives from the media and a
broader scientifically interested public is the result of the entertaining
fictional qualities of its "revolutionary" achievements.

This developments take place against the background of frustration within the
particle physics community as a result of inconclusive or failed attempts to
make further progress with the standard model (SM). The latter has remained
particle physics finest achievement ever since its discovery more than three
decades ago. It continued the line of gradual unification which started
already with Faraday, Maxwell, Einstein Heisenberg and others. This kind of
unification has been the result of a natural process of the development of
ideas, i.e. the protagonists did not set out with the proclaimed aim to
construct a TOE, rather the coming together was the result of a natural
unfolding of ideas following the intrinsic theoretical logic of principles,
but always with observations and experiments having the ultimate say.

Particle physics is a conceptually and mathematically quite demanding science
and its progress sometimes requires an adventurous\ "into the blue yonder"
spirit. But precisely because of this it needs a critical balance whose
intellectual depth at least matches that of its target. To some extend this is
part of an inner theoretical process in which the main issue is that of
conceptual consistency. Particle theory is very rich in established
fundamental principles, and a good part of theoretical research consists in
unfolding their strong intrinsic logic in the concrete context of models.
Experiments cannot decide whether a theoretical proposal is conceptually
consistent, but they can support or reject a theory or select between several
consistent theories and last not least lead to a limitation of the validity of
principles and hints how to amend them.

A theoretician should carry his criticism, if possible, right into the
conceptual-mathematical core of a theory. The most basic property in the
formulation and interpretation of models in particle physics has been and
still is the \textit{localization of objects in spacetime}. With respect to
string theory this amounts to the question whether it really represents, as
its name indicates, a theory of string-like localized objects in spacetime.
This will be the main theme in section 6. It is not easy to criticise
something which is conceptually and mathematically as opaque as string theory.
Fortunately this question allows a rigorous answer in the case of the
"founding" model of string theory: the canonically quantized Nambu-Goto model
\cite{N}\cite{G}.

The answer will be that the quantum string is described by a generalized free
field whose infinite mass/spin tower spectrum is such that the relative
weights of irreducible components in the Kallen-Lehmann function depend on the
chosen classical string configuration. In other words the possible
configurations associated with the classical N-G functional integrand become
encoded into a mass/spin tower of a point-like infinite component field but
have no bearing on spacetime localization i.e. on "wiggling strings" in
spacetime. This calls for a more general understanding of the relation between
classical and quantum localization beyond the point-like case (where both
coalesce) based on an autonomous notion quantum localization.

String theory was conceived as an S-matrix theory. An S-matrix theory has no
direct relation to localization, the indirect link goes only through the
inverse scattering problem. Two previous attempts at a pure S-matrix theory,
that of Heisenberg's 1943 and the Chew-Mandelstam-Stapp bootstrap idea of the
60s failed, because contrary to naive expectation of uniqueness, at the end of
the day there were too many solutions (the infinite family of 2-dim.
factorizing models). The dual resonance model on the other hand, which
resulted from replacing the general crossing property by the
Veneziano--Virasoro-Dolen-Horn-Schmid duality, led to a very explicit formulae
which represented some (only at that time desirable) properties of an S-matrix
describing strong interactions, but had serious conceptual problems.

A critical presentation of string theory would not be comprehensive if it
ignores the history which led to it. The best way to try to understand a
theory which derives its computational recipes from unclear physical concepts
is to critically follow the path of its history. This is precisely what we
will do in the following four sections.

The present view of string theorists at their own history is reflected in the
several recent 40 year anniversary contributions, some from experienced
veterans of the dual model days with first hand insider knowledge. They make
quite interesting historical reading and some of their points will be
mentioned in later sections. But they were written in the certainty of ideas
which at the end led to a theory of great impact and world wide dominance
which is certainly not supported by the point of view in this essay.

The criticism of the scientific aspects of superstring theory and its
sociological manifestation in this essay is not directed against its
protagonists and individual contributors. Rather its main target is the lack
of balance caused by the uncritical reception of its central claim to
represent a quantum theory in which string-like objects replace the standard
point-like fields of QFT. It will be argued that its unreal dreamlike almost
surreal physical appearance which it has outside the string community, is the
result of a mismatch between what it claims to be and what it really is. More
specifically, the geometric aspects which led to its name are not identical
with its intrinsic physical localization property.

For a mathematician this is irrelevant, the geometric properties should be
present somewhere, but he could not care less about where it is encoded,
whether the mathematics of string configurations describes matter localized in
spacetime or whether it becomes encoded into an infinite component field which
"sits over one point". A physicist however is required to follow the logic of
quantum localization; if he would base his interpretation solely on the
geometric logic and overlook that localization has an intrinsic quantum
physical meaning, he would go astray or, in the terminology of this article he
would take the path of wrong metaphors and not that of autonomous physical
interpretation. The historical context shows that the erosion of the intrinsic
by the metaphoric, of which string theory is the most visible illustration,
did not come out of the blue.

Even at the risk of sounding cynical there is also some positive aspect,
namely the construction of the free Nambu-Goto string or the superstring
solves a problem which a whole previous generation (Barut, Kleinert, Fonsdal,
Ruegg, Budini,.. ) found a hard nut to crack namely to find natural quantum
constructions of infinite component wave functions or (which is the same)
infinite component free fieds. The quantum mechanics archetype was the
hydrogen atom which can be fully understood in terms of a representation of
the noncompact group SO(4,2). Attempts to get a relativistic mass/spin
spectrum along these line failed. If one is not prudish about spacetime
dimensions, then the N-G or superstring is indeed a solution. The secret is to
replace group theory with vector-valued (or dotted/undotted spinorial) quantum
mechanical oscillators on which the Lorentz group acts. String theory uses
oscillators as they arise from the Fourier decomposition of multicomponent
currents of a chiral QFT. But beware, this does not mean that one embeds a
conformal field theory as a stringy one-dimensional subalgebra in higher
dimensional spacetine. All these problems wil be addressed in detail in later sections.

It is an interesting question whether the problems caused by confusing
geometric properties with the intrinsic meaning of localization have a higher
dimensional analog beyond the one-dimensional strings. The lack of a pure
quantum (non quasiclassical) brane solution in analogy to the N-G model
prevents presently a direct answer to this question but the negative
experience with string theory would go against describing branes by embedding
in terms of quantized coordinates which depend on more parameters as for strings.

This intention to go to the \textit{conceptual} roots separates the content of
the present essay from several other books and articles which have appeared
\cite{Wo}\cite{Sm}\cite{Hedrich}. We want to go beyond the mere assertion that
string theory has surreal aspects; it is our intentions to expose the cause of
its misleading metaphor. We do however agree with all the authors who have
written string-critical articles that even if superstring theory would be
consistent on the conceptual-mathematical level, the lack of tangible results
despite of more than four decades of hard work by hundreds of brilliant minds,
the futile consummation of valuable resources and last not least its bizarre
philosophical implications should be matters of great concern.

The crisis in particle physics on the eve of the LHC experiments has strong
relations to a ongoing crisis of post cold war socioeconomic system and its
ideology. The idea of a final theory of everything is too close to the way
globalized capitalism views itself in order to resist the temptation of
looking in more detail on some surprising parallels. As will be argued in the
last section both developments are characteristic manifestations of the same
Zeitgeist. It remains to be seen whether the turbulent 2008 crash of the post
cold war economic system and its ideological of "end of history" setting finds
its analog in a loss of support for project of a TOE and the superstring metaphors.

The content is structured as follows.

The next section reviews Heisenberg's S-matrix proposal and its profound
criticism by Stueckelberg on the basis of macro-causality problems. The third
section recalls the S-matrix bootstrap program whose lasting merit consists in
having added the important on-shell crossing symmetry to the requirements of
an S-matrix program. The fourth section analyses the relation of on-shell
crossing property with off-shell localization concepts and comments on its
proximity to the Kubo-Martin-Schwinger (KMS) thermal aspect of localization.
Section 5 reviews the implementation of duality of the Dolen-Horn-Schmidt
(DHS) dual resonance model in the setting of a multi-charge chiral current
mode. In this way the differences between duality resulting from the
generalized (anyonic) commutation relations of charge-carrying chiral fields
and the particle-based notion of crossing becomes highlighted. The prior
results on quantum localization obtained in the fourth section are then used
in section 6 to show that string theory of the canonical quantized Nambu-Goto
model, contrary to terminology, does not deal with string-localized objects in
spacetime. Rather the classical string configurations associated with the
functional integrand become encoded into the structure of and infinite
component local field whose decomposition leads to free fields corresponding
to a discrete mass/spin tower.

The last section is an attempt to shed some light on how a theory with so many
conceptual shortcomings as well as lack of predictive power was able to
represents the spirit of particle physics at the turn of the millennium. In
that section we leave the ivory tower of particle theory and turn to some
observations on parallels between between superstring physics and the
millennium Zeitgeist.

Since the mathematical-conceptual content is quite demanding and we want to
keep this essay accessible to readers with more modest mathematical knowledge,
some statements and arguments will appear more than once in a different
formulation and context.

\section{QFT versus a pure S-matrix approach from a historical perspective}

Particle physics was, apart from a period of doubts and confusion around the
ultraviolet catastrophe which started in the late 30s and did not last longer
than one decade, a continuous success story all the way from its inception
\cite{Dar} by Pascual Jordan (quantization of wave fields for light waves and
matter) and Paul Dirac (relativistic particles and anti-particles via hole
theory) up to the discovery of the SM at the end of the 60s.. For about 40
years the original setting of Lagrangian quantization, in terms of which QFT
was discovered, gave an ever increasing wealth of results \textit{without
requiring any change of the underlying principles}. After the laws of QT had
been adapted to the changed conceptual and mathematical requirements resulting
from the causal propagation in theories with a background-independent maximal
velocity (the velocity of light), the new principles were in place and the
subsequence significant progress, including the elaboration of renormalized
QED, consisted basically in finding new conceptual-mathematical realizations
of those principles underlying QFT. The ultraviolet divergencies posed some
temporary obstacle on this path, since they led to the proposal of an
elementary length which would have required a modification of the
principles\footnote{Despite of numerous attempts, some of them even ongoing,
it has not been possible to find a conceptual framework for nonlocal theories.
The notion of a cutoff has remained a metaphor. Non of the soluable
two-dimensional local factorizable models permits the introduction of a
cutoff, without wrecking its physical content and loosing the mathematical
control.}. After the clouds of doubts about the ultraviolet catastrophe
dispersed, thanks to the new setting of covariantly formulated perturbative
renormalization theory, the conceptional and mathematical improvements
reinforced the original principles.

It is interesting to observe that, already at the beginning of QFT, its
protagonist Pascual Jordan worried about the range of validity of
quantization. His doubts originated from his conviction that, although
classical analogies allow in many cases rapid access to the new quantum theory
of fields in form of important perturbative model illustrations, in the long
run a more fundamental quantum theory should not need the quantization
parallelism to the less fundamental \textit{classical} Lagrangian formalism.
Rather it should develop its own autonomous tools for the classification and
construction of QFTs, or in his words "without classical crutches" \cite{Kha}.
To turn the argument around: to the extend to which one has to still rely on
quantization crutches, one has not really reached the conceptual core of the
new theory.Jordan's doubts about the range of validity of that umbilical cord
to classical field theory did not originate from any perceived concrete
shortcoming of his "quantum theory of wave fields". Rather the state of
affairs in which he discovered this new theory did not comply with his
philosophical conception; in his opinion a classical parallelism can only be
tolerated as a temporary device for a quick exploration of those parts of the
new theory which are in the range of this quantization recipe.

But things did not develop in the direction of his plea. The ultraviolet
divergence crisis of the 30s ended in the late 40s in the discovery of
renormalized QED, a fact which certainly revitalized the Lagrangian approach
and pushed the search for an intrinsic formulation into the sideline.

Unfortunately the renormalized perturbation series of quantum field
theoretical models diverges, so the hope to settle also the existence problem
of QFTs in the Lagrangian quantization setting did not materialize; the
success of the renormalized perturbative setting did not lead to a conceptual
closure of QFT. However at least it became clear that the old problem of
ultraviolet infinities, which almost derailed the development of QFT, was in
part a pseudo-problem caused by the unreflective use of quantum mechanical
operator techniques for pointlike quantum fields which are too singular to
qualify as operators.

The main difference of the status of QFT to any other physical theory
(statistical mechanics, quantum mechanics,..) is that as a result of many
interesting nontrivial models with solid mathematically control, one knows
that the "axioms" which characterize them admit nontrivial solutions. It is
often forgotten or pushed aside as a nuisance, that \textit{QFT has not yet
reached this state}, it still has a long way to go.

Using more adequate mathematical tools in conjunction with a minimality
principle which limits the short distance singularity in every perturbative
order \cite{E-G}, one finds that there are local couplings between pointlike
fields for which the perturbative iteration either does not require more
parameters than there were in the beginning, or adds only a finite number of
new couplings which one could have already included in the starting
interaction expressed in terms of Wick-product of free fields\footnote{I am
referring here to the Epstein Glaser \cite{E-G} formulation which produces the
renormalized finite result directly by treating the fields in every order
according correctly according to their singular nature. The avoidance of
intermediate cutoffs or regularizations maintains the connection with the
quantum theoretical Hilbert space structure of QFT.}. The renormalized theory
forms a finite parameter space on which the (Petermann-Stueckelberg)
renormalization group acts ergodically. These finite parametric families are
conveniently pictured as "islands" in an infinite parameter setting (the
Bogoliubov spacetime dependent operator-valued S-functional, or the Wilson
universal renormalization group setting) within a universal master
S-functional\footnote{The S-functional is a formally unitary off-shell
operator which contains the on-shell S-matrix. The existence of the latter
does however not depend on that of the former.} depending on infinitely many
coupling functions (which by itself has no predictive power). Since the
renormalization group leads from any point on the island in coupling space to
any other point on the same island, a QFT cannot provide a method to
distinguish special numerical values as that e.g. of the fine-structure
constant.. \ 

The phenomenon of interaction-caused infinite vacuum polarization clouds
(finite in every order perturbation theory) gives rise to a conceptual rupture
with QM \cite{interface} and leads to a change of parameters in every order.
But since these parameters remain undetermined anyhow, this causes no harm.
The inexorable presence of interaction-induced vacuum polarization prevents
one to think of an initial numerical (Lagrangian) value for these parameters
which is then changed by a computable finite amount. With other words unlike
in QM there is no \ separate "bare" and "induced" part. This is why the
Epstein-Glaser renormalization, which avoids such quantum mechanical concepts,
is conceptually preferable \cite{E-G}. It not only addresses the singular
nature of fields but it also exposes the limits of QFT concerning the
predictive power about the numerical value of certain parameters in a more
honest way. \ 

So when string theorists say that their theory is ultraviolet finite and
contrast this with QFT, what they really mean in intrinsic terms is that their
theory is more economical (and hence more fundamental) in that it has only the
parameters which describe string interactions i.e. the string tension. But
beware, they say that without being able to give a proof!

This implies that in particular that \textit{string theory has no vacuum
polarization} which is of course completely consistent with its on-shell
S-matrix character. An S-matrix is free of vacuum polarization par excellence,
in fact in QFT the S-matrix and formfactors are the only objects of this kind.
Heisenberg's plea for basing particle physics on the S-matrix was proposed
precisely because of this absence of vacuum polarization and the ensuing
ultraviolet problems. But can one really do particle physics without such a
central concept as vacuum polarization? can one formulate physically motivated
autonomous S-matrix properties without the intervention of fields ? These are
clearly the kind of problems one faces in a pure S-matrix approach. We will
return to them in the historical context of Stueckelberg and Heisenberg.

Before taking a critical look at pure S-matrix attempts, it is quite
instructive to glance at some unfinished problems of QFT.

The well-known power counting restriction to interactions $dim\mathcal{L}%
_{int}=$ $d$ (spacetime dimension)\footnote{The causal perturbative approach
(in contrast to the functional integral setting) does not use the Lagrangian
formalism; $\mathcal{L}_{int}$ is only a notational device for the
Wick-ordered polynomial which represents the pointlike interaction between
fields.} is quite severe; in $d=4$ it only allows pointlike fields $\Phi$ with
short distance dimension $sdd\Phi=1.$ In addition massless vectorpotentials
(and more generally the potentials associated with the Wigner massless finite
helicity representations) \textit{cannot be pointlike covariant objects within
a (ghost-free) Wigner Fock representation}; the best one can do is to permit a
spatial semiinfinite extension. In the case of (m=0,s=1) this leads to
semiinfinite stringlike-localized covariant vector potentials $A_{\mu}(x,e)$
\cite{MSY}\ with $sdd=1~$which act in the same physical Wigner-Fock space as
the associated pointlike field strength; in fact their only unphysical aspect
is that they are not \textit{local} observables. As expected their stringlike
localization shows up in the commutator between two potentials whenever the
semiinifinite string of one $x+\mathbb{R}_{+}e$ gets into the influence region
of the other. Such string-localized potentials with $sdd=1$ independent of
helicity exists for all $(m=0,s\geq1)\footnote{For s+2 the field strength is
the (linearized) Riemann tensor and the potential is a string-localized
linearized metric tensor $g_{\mu\nu}(x,e)$ localized along the line
$x+\mathbb{R}_{+}e$ (see section 4). The string localization comes from
quantum requirements and has no counterpart in the classical theory.}.$

For massive fields there is no such representation theoretic reason to
introduce string-localized fields since pointlike covariant fields exist for
all admissible covariant spinorial objects $\Phi^{(A,\dot{B})}$ with
$\left\vert A-\dot{B}\right\vert \leq s\leq A+\dot{B}$. But it is well-known
that the short distance dimension increases with the spin s; for $s=1$ it is
at least $sdd=2.$ They violate the power counting theorem and therefore the
lowering of the dimension to $sdd=1$ for semiinfinite stringlike localized
fields \cite{MSY} is of potentially great interest for extending the realm of
renormalizable interactions. The crucial question is whether there exists a
perturbation theory for stringlike localization, more specifically whether the
Epstein-Glaser iteration step, which is build around pointlike locality,
admits an extension to semiinfinite stringlike localization. This is presently
under investigation and we refer to work in progress \cite{M-S}.

Another way to lower short distance dimensions but without sacrificing the
pointlike formalism is to allow \textit{indefinite metric} with the help of a
pointlike (BRST) ghost formalism. The use of such a formalism is suggested by
the quantization of classical gauge theories. The requirement of gauge
invariance is the key which permits to return to a Hilbert space. The
formalism is quite efficient but it is primarily directed to the perturbative
construction of local observables which are identified with the gauge
invariant objects. Nonlocal physical operators as charged fields have to be
defined "by hand" \cite{infra} which up to now has only been possible in
abelian gauge theories. In an approach based on string-localized fields the
nonlocality would be there from the beginning and all objects local and
nonlocal would be part of the same formalism.

In the massive case the stringlike formulation would permit to start already
with $sdd=1$ massive vectormeson and one would have the chance to understand
better if and why the requirement that the string localization should only be
a dimension-lowering technical trick (i.e. the theory continues to be
pointlike generated) requires the presence of another scalar scalar field
(naturally with a vanishing one-point function).

This short account of the history of QFT and particle physics contains most of
the ideas which are needed for the formulation of the SM which places QED, the
weak interaction and the QCD setting of strong interactions under one common
gauge theoretic roof. But it also was meant to expose some important
unfinished areas of QFT. Once one goes beyond textbook presentations of QFT
and returns to the principles, QFT is to a large extend virgin territory. The
claim that QFT is a closed subject and that its innovative role has passed to
string theory may be is not really an argument against QFT but rather against
the carricature picture wgich string theorists created about QFT.

One of the marvelous achievements of the post QED renormalization theory is a
clear understanding of the particle-field relation (not to be confused with
the particle-wave dualism in QM) in the presence of interactions. Whereas in
free field theories Heisenberg had already observed the presence of vacuum
fluctuations due to particle-anti-particle pairs in states obtained by the
application of (Wick) composites to the vacuum, the real surprise came when
Furry and Oppenheimer discovered that in interacting theories even the
Lagrangian "elementary" field generates vacuum polarization upon application
to the vacuum state. Different from the case studied by Heisenberg, the
interaction-induced polrization pairs increase in number with the perturbative
order and form a \textit{vacuum polarization cloud} containing an infinite
number of virtual particles. This observation challenges the naive
identification of particles and \ fields which is the result of a
simple-minded conceptual identification of QFT as a kind of relativistic QM.
Although one-particle states exist in the Hilbert space and the global
operator algebra certainly contains particle creation/annihilation operators,
\textit{compactly localized subalgebras} in interacting QFTs contain no
vacuum-\textbf{p}olarization-\textbf{f}ree \textbf{g}enerator
(PFGs)\textit{\footnote{The only localization which allows PFGs is the
non-compact wedge-like localization \cite{BBS}.}.} In other words in the
presence of interactions (independent of what kind of interaction) there are
no operators localized in subwedge regions which creates a one-particle state
from the vacuum without being accompanied by an infinite vacuum polarization cloud.

The at that time mysterious particle/field relation was partially unveiled
when in the post QED renormalization period it became clear that interacting
\textit{QFT is not capable to describe particles at a finite time}; as a
result of the ubiquitous presence of vacuum polarization clouds it is only
possible to have an asymptotic particle description when, barring long range
forces and infrared problems, the localization centers of particle are far
removed from each other, so that the interaction is effectively switched off.
In fact the elaboration of scattering theory as a structural consequence of
causal locality, energy-momentum positivity and the presence of a mass gap, as
carried out in the late 50s and early 60s, was one of the finest achievements
of relativistic particle theory. No comparable conceptual enrichment has been
added after the discovery of the SM.

As mentioned in the previous section, the idea of a pure S-matrix theory as a
remedy against the ultraviolet catastrophe of the old (pre-renormalization)
QFT was first proposed by Heisenberg\footnote{The concept of a unitary
scattering operator as a mapping incoming multiparticle configurations into
outgoing in the limit of infinite timelike separations was introduced
independently by Wheeler and Heisenberg. however the idea of a pure S-matrix
theory as an antidote against the pre-renormalization pretended ultraviolet
catastrophe is attributed to Heisenberg.} \cite{1946}. The S-matrix models
with which he illustrates his ideas resulted from a naive unitarization of the
interaction Lagrangian (see next section). Heisenberg's proposal was
immediately criticized by Stueckelberg who pointed out that, although it was
Poincar\'{e}-invariant and unitary, it did not meet the requirements of macro-causality.

In the next section we will comment on Heisenberg's construction and isolate
the problem on which all pure S-matrix theory failed: fitting together
unitarity and Poincar\'{e} covariance with macrocausality (notably the cluster
factorization property). Clustering is the spacelike aspect of macro-causality
which is indispensable for any S-matrix, whether its comes from QFT or any
other theory of interacting particles. In QFT and other off-shell
implementations of particle interactions, the clustering property is
implemented on correlation functions or (similar to nonrelativistic QM)
through asymptotic additivity of the interaction-dependent generators of the
Poincar\'{e} group. Its validity for the asymptotic configurations is then a
side result of the proof of asymptotic convergence. With other words, the
highly nonlinear on-shell unitarity requirement is trivialized by showing that
it results from the large time limiting of more easily implementable linear
additive clustering properties for correlation functions.

A long time after the Heisenberg S-matrix proposal was abandoned and QFT
experienced a strong return in the form of renormalized quantum
electrodynamic, ideas about an S-matrix-based approach returned, this time as
the result of the failure of perturbative arguments in strong interactions
between mesons and nucleons. This led to the \textit{S-matrix bootstrap} by
Chew and Mandelstam. The ideological fallout of the bootstrap approach
interspersed with ideas from eastern mysticism entered the popular writing of
the physicist F. Capra and others. The bootstrap ideas never led to any
tangible remaining results in particle physics, but their influence on the
popular "new age" culture and the Zeitgeist of the 60s and 70s has been considerable.

The analytic aspects of QFT correlations, which follow from locality and
spectral properties, imply a kind of crossing relation which was first seen in
Feynman diagrams within a fixed perturbative order. Restricting the external
legs of these graphs to the mass-shell in order to obtain perturbative
contribution to the S-matrix, one was able to show that the different S-matrix
elements belonging to different distributions of n-particles into $k$ incoming
and $l$ outgoing particles are connected by an analytic continuation. The
surprising aspect (which was not trivial even with Feynman graphs) was that
this was possible without leaving the complexified mass shell. With other
words crossing is not a symmetry but rather an analytic on-shell mark left by
the spacelike commutativity of QFT. Although there is presently no general
proof of crossing for generic particle configuration beyond some special
cases, most particle physicists would agree that highlighting this important
property will remain as one of the few legacies of that S-matrix bootstrap period.

The S-matrix bootstrap community did not succeed to come up with a model in
which this new property is nonperturbatively realized. Strictly speaking a TOE
has no model representation which is different from itself; it leads to an
"everything or nothing" alternative.\ Whereas crossing and unitarity were the
two main postulates in the S-matrix bootstrap setting, other important
properties as Stueckelberg's macrocausality (in particular the cluster
factorization) were missing from the bootstrap postulates.

There exists an exceptional situation in d=1+1\footnote{This is related to the
kinematical equality of the energy-momentum delta function with the product of
two one-particle delta functions which only holds in d=1+1.}. This has the
effect that the cluster property cannot separate the genuine 2-particle
interacting contribution from the identity term of S. In this case it is
possible to fulfill unitarity and crossing with purely elastic two-particle
S-matrices. In fact one can classify such two-particle S-matrices and show
that all higher elastic processes are given by a combinatorial formula in
terms of the two-particle S-matrix \cite{Ba-Ka} Purely elastic relativistic
scattering in higher spacetime dimension, as it occurs in the relativistic
quantum mechanics of \textit{direct particle interactions} \cite{Co-Po} (see
next section), is not possible in QFT.

\section{Unitarity and macro-causality in relativistic particle theories}

There are three fundamental requirements which an S-matrix of relativistic
particle physics in must obey, namely Poincar\'{e} invariance, unitarity and
macro-causality. None of these concepts requires to introduce fields;
macrocausality is a very weak version of causality which can be formulated and
understood in terms of only particle concepts. To avoid misunderstandings,
there are analytic properties of scattering amplitudes as, e.g. the crossing
property which requires analytic continuation inside the complex mass shell.
Such analytic on-shell properties cannot be traced back to principles
referring to particles only; rather they must be understood as being an
on-shell imprint of the causal locality principles of an underlying local
quantum physics. i.e. they are consequences of the existence of local fields
which interpolate between the incoming/outgoing particles.

As a pedagogical exercise, which leads right into the problematic aspects of
pure S-matrix theories, let us revisit the situation at the time when
Stueckelberg \cite{Stue} criticized Heisenberg's S-matrix proposal.

As already mentioned, Heisenberg suggested that avoiding the vacuum
polarization in interacting QFTs by abandoning fields in favor of directly
constructing S-matrices could lead to a solution of the ultraviolet problem.
His rather concrete proposals consisted in expressing the unitary S-matrix in
terms of a Hermitian "phase operator" $\eta$ and imposing physically motivated
restrictions on this operator$.$ In modern notation his proposal reads%

\begin{align}
S  &  =\exp i\eta\\
\eta &  =%
{\displaystyle\sum}
\frac{1}{n!}%
{\displaystyle\int}
...%
{\displaystyle\int}
\eta(x_{1},...x_{n}):A_{in}(x_{1})...A_{in}(x_{n}):dx_{1}...dx_{n}\nonumber\\
&  \eta_{Hei}=g\int:A_{in}^{4}(x):d^{4}x\nonumber
\end{align}
where the on-shell coefficient functions of $\eta$ are chosen to be
Poincar\'{e} invariant and subject to further physically motivated
restrictions. In fact one such restriction which he suggested was that the
on-shell $\eta$ should be close to a Lagrangian interaction i.e. have local
coefficient functions as illustrated in the third line. It is customary to
split off the identity operator from $S$ and formulate unitarity in terms of a
quadratic relation for the T-operator
\begin{align}
&  S=1+iT\\
&  iT-iT^{\ast}=TT^{\ast}\nonumber
\end{align}
In this form the unitarity is close to the optical theorem and convenient for
perturbative checks.

Unitarity and Poincar\'{e} invariance are evidently satisfied if the (possibly
singular) functions $\eta(x_{1}...x_{n})$ are Poincar\'{e} invariant, but what
about macro-causality? For spacelike separation one must require the so called
\textit{cluster factorization property}. If there are n+m particles involved
in the scattering (the sum of incoming and outgoing particles) and one forms k
clusters (again containing in and out) and then separates these clusters by
large spacelike translations, the S-matrix must factorize into the product of
k smaller cluster S-matrices referring each describing the scattering
associated with a cluster. For the simplest case of two clusters
\begin{equation}
\lim_{a\rightarrow\infty}\left\langle g_{1}^{a},..,g_{m}\left\vert
S\right\vert f_{1}^{a},..,f_{n}\right\rangle =\left\langle g_{1},..\left\vert
S\right\vert f_{1},..\right\rangle \cdot\left\langle ..g_{m}\left\vert
S\right\vert ..f_{n}\right\rangle
\end{equation}
where the first factor contains all the a-translated wave packets i.e. the
particles in the first cluster and the second factor contains the remaining
wave packets. \ In massive theories the cluster factorization is rapidly
attained. \ This asymptotic factorization property is usually written in
momentum space as
\begin{equation}
\left\langle q_{1},..q_{m}\left\vert S\right\vert p_{1},..p_{n}\right\rangle
=\delta-contrib.+~products~of~lower~delta\text{ }contrib.
\end{equation}
i.e. the S-matrix contains besides the \textit{connected} contribution the
disconnected parts which consists of products of connected amplitudes
referring to processes with a lesser number of particles. The connected parts
have the correct smoothness properties as to make the formulas meaningful.

For timelike separated clusters the fall-off properties for large cluster
separations are much weaker. In fact there are inverse power law corrections
in the asymptotic timelike cluster distance. With the correct $i\varepsilon$
prescription in momentum space they define what is referred to as
\textit{causal} re-scattering or\textit{ the causal one-particle structure};
the presence of this singularity structure prevents the presence of time-like
precursors\footnote{An example for a model which was shown \cite{Swieca} to
lead to such timelike precursors (as the result of the presence of complex
\ poles) was the Lee-Wick model.}.

For an explanation imagine a kinematical situation of elastic 3-particle
scattering in which the third particle enters the future cone of an
interaction region of particle 1 and 2 \ a long time after the 1-2 interaction
happened, and then scatters with the outgoing first particle leaving particle
2 undisturbed. \ In the limit of \textit{infinite timelike separation} the
connecting trajectory which describes the path of particle 1 between the 1-2
interaction region and the later 1-3 interaction region must coalesce in the
limit with a causal propagator. i.e. asymptotically this 3-particle scattering
must contain a singular Feynman propagator connecting two 2-particle
scattering processes.

Whereas the cluster factorization of a Heisenberg S-matrix Ansatz is a trivial
consequence of imposing the connectedness property on the coefficient
functions of the phase operator $\eta$, it is not possible to satisfy the
causal one-particle structure with a finite number of terms in $\eta;$ in fact
no pure S-matrix scheme has ever been devised which secures the validity of
the causal one-particle structure in the presence of unitarity.

At this point the weakness of a pure S-matrix approach as advocated by
Heisenberg becomes exposed; although one can formulate all three requirements
solely within a particle setting, one lacks an implementing formalism. The
phase matrix $\eta$ which linearises unitarity unfortunately complicates those
properties which were linear in terms of $S.~$To construct solutions by
"tinkering" for objects which besides linear properties also obey unitarity,
has throughout the history of pure S-matrix attempts never led anywhere .

It is off-shell QFT and its asymptotic timelike convergence properties, better
known as scattering theory, which saves us for spending the rest of our days
with S-matrix tinkering. The QFT correlation functions are the natural arena
for implementing causality properties; the observables are Hermitian and not
unitary and the building up of S-matrix unitarity is part of the asymptotic
convergence whose existence is guarantied by the properties of the correlations.

This problem of causal re-scattering in a Heisenberg S-matrix setting, and
more generally in any pure S-matrix formulation, was what finally convinced
Stueckelberg \cite{Stue} that a pure S-matrix approach is not feasible.

The S-matrix is without doubt the most important observable concept in
particle physics, but it should remain the "crown" of the theory and not its
foundation nor its principal computational tool. This was at least the gist of
Stueckelberg's critique on Heisenberg's program when he pointed out that to
reconcile macro-causality with unitarity "by hand" (i.e. without an off-shell
setting which naturally unites these seemingly ill-fitting on-shell concepts)
one runs into insoluble problems.

Interestingly enough, Stueckelberg then combined his idea of the causal one
particle structure with postulating pointlike interaction vertices leaving out
unitarity and in this way came to Feynman rules several years before Feynman,
but without knowing that he arrived at the perturbative rules of a QFT. For
showing that this prescription leads to on-shell unitarity, at least on a
perturbative level, he lacked the elegance of the formalism of QFT in which
the on-shell unitarity (and all the other properties of S) is derived from
simpler properties of correlation functions.

A systematic step for step derivation from a covariant Tomonaga setting of
QFT, including the Schwinger or Feynman formalism of renormalization, and with
particular care concerning the perturbative connection between QFT and the
S-matrix, was finally given by Dyson. It was also Dyson who raised the first
doubts about the convergence of the renormalized perturbative series.

The conceptually opaque status of perturbation theory lends importance to a
purely structural derivations of particles properties and scattering data
directly from the quantum field theoretic principles. Without having
mathematically controlled models at one's disposal, structural arguments
became increasingly important. Despite all the difficulties to construct
interacting models there was no problem to define the requirements which are
characteristic for QFT in mathematical clear terms. This "axiomatic setting"
in terms of correlation functions of products of fields (Wightman functions)
was a major achievement. Under the assumption of a mass-gap in the energy
momentum spectrum it led to the validity of (LSZ) scattering theory; in this
way it became a framework which combined particles with fields. It led to
systems of nonlinear equations for time-ordered or retarded functions whose
perturbative solutions contained those obtained from the Lagrangian
quantization formalism. One of its nonperturbative results was the derivation
of Kramers-Kronig type of dispersion relation and their experimental
verification. It was later refered to as "axiomatic QFT", but at least its
original motivation was driven by the pragmatic desire to go beyond divergent
perturbative series.

The derivation of the Kramers-Kronig dispersion relations for the scattering
amplitudes in particle physics and its subsequent experimental verification is
an example of particle theory at its best, it secured the localization
properties of QFT up to present energies and it did so in a clear direct and
yet modest way without relying on metaphors or importing
geometric-mathematical ideas beyond those which are autonomous to QFT. In
comparison to ST it was one of particle physics finest achievements.

All this was achieved less than a decade after Stueckelberg's criticism of a
\textit{pure} S-matrix approach and the discovery of renormalized perturbation
theory by Tomonaga, Schwinger, Feynman and Dyson and forms the backbone of the
LSZ and Haag-Ruelle scattering theory.

As indicated above, the basic simplification of the problem of macro-causality
for the S-matrix consisted in the realization that its representation as the
large time scattering limit \textit{defuses the rather intractable nonlinear
problem} \textit{of implementing macro-causality in the presence of unitarity}
by delegating it to simpler linear (off-shell) properties for correlation
functions. The path from local observables to the S-matrix is generally not
invertible. In a QFT in which all formfactors (matrix elements between bra in-
and ket out-states) including the S-matrix (the formfactor of the identity
operator) fulfill the crossing property, the inversion turns out to be
unique\footnote{Such inverse scattering problems show very clearly the
conceptual advantage of formulating QFT in terms of spacetime-indexed nets of
algebras rather than in terms of pointlike field coordinatizations of the
Lagrangian quantization. The crossing symmetric S-matrix is not capable to
highlight individual field coordinatizations, it only fixes the local net.}
\cite{unique}. Within the family of two-dim. factorizing S-matrices the
existence of the associated QFT can actually be proven \cite{Lech}..Whether
the general framework of QFT can also lead to a nonperturbative classification
and construction of higher dimensional QFT remains to be seen.

There is another quantum mechanical particle physics setting in which a
Poincar\'{e} invariant unitary macro-causal S-matrix arises through scattering
theory in the large time asymptotic limit: \textit{Direct Particle
Interaction} (DPI). It forgoes micro-causality and fields and only retains
Poincar\'{e} covariance and macro-causality. It is certainly more
phenomenological than QFT since it contains interaction potentials instead of
coupling strength.

The reason why it is mentioned here (even though we are not advocating its use
outside medium energy pion-nucleon physics) is because \textit{its very
existence} not only \textit{removes some prejudices and incorrect folklore}
(including the belief that relativistic particle interactions are necessarily
QFTs or that a clustering S-matrix matrix can only arise from a QFT setting),
but it also indicates what has to be added/changed in order to arrive from
particle interactions to a full QFT setting.

Relativistic QM of particles is based on the Born-Newton-Wigner
localization\footnote{This terminology refers to the Born probability of the
associated with relativisic wave function. As pointed out by Newton and
Wigner, the relativistic normalization leads to a change as compared to the
nonrelativitic Schroedinger QM.}, whereas the causal localization of QFT,
which incorporated the finiteness of the propagation speed, is related to the
Poincar\'{e} representation theory via modular theory (next section). The
B-N-W localization of wave packets is sufficient for recovering the forward
lightcone restriction for 4-momenta associated with events which are separated
by large time-like distances. Although this suffices to obtain a Poincar\'{e}
invariant macro-causal S-matrix, it fails on securing the existence of local
observables and vacuum polarization. For a presentation of the differences and
their profound consequences see \cite{interface}.

This DPI scheme introduces interactions between particles within a
multiparticle Wigner representation-theoretical setting by generalizing the
Bakamijan-Thomas (B-T) two-particle interacting Poincar\'{e} generators
\cite{Co-Po}. But whereas in the nonrelativistic QM\ the additivity of the
interaction potentials trivializes the problem of cluster factorization, there
is now no such easy connection between the modification of the n-particle
Poincar\'{e} generators and the cluster properties of the interactions.
Nevertheless, by using the notion of \textit{scattering equivalences} one can
arrive at a cluster factorization formula for the interacting Poincar\'{e}
generators and the and the S-matrix \cite{Co-Po}\cite{interface}. A scattering
equivalence consists in a unitary transformation which changes the
representation of the Poincar\'{e} generators but maintains the S-matrix. In
the Coester-Polyzou DPI scheme the iteratively defined (according to particle
number n) Poincar\'{e} generators lack the large distance additivity
associated with clustering, but a scattering equivalence transformation
rectifies this situation.

One starts with a B-T two-particle interaction and computes the 2-particle
Moeller operator and the associated S-matrix as a large time limit of
propagation operators. As in the nonrelativistic case the two-particle cluster
property is satisfied for short range two particle interactions. For 3 and
more particles the construction of cluster factorizing Poincar\'{e} generators
and S-matrices require the iterative application of scattering equivalences.
The so constructed 3-particle S-matrix clusters with respect to the 2 particle
S-matrix in the previous step. But in contrast with the nonrelativistic
situation it also contains a 3-particle connected part which vanishes if any
one of the particles is removed to spacelike infinity and their is no natural
restriction to only two-particle interactions: in other words the occurrence
of direct higher particle induced interactions cannot be prevented in any
natural way.

As a result of the use of scattering equivalences in order to achieve
clustering, there is no natural way to encode such multiparticle theories into
a second quantization Fock formalism. They are basically relativistic S-matrix
theories because their only truly covariant object is the Poincar\'{e}
invariant S-matrix. In particular the DPI setting does not lead to covariant
formfactors. In the original formulation of DPI the scattering was purely
elastic, but later it was shown that an extension with particle creation
channels is possible. Hence the characteristic difference of DPI to QFT is not
the presence of creation/annihilation channels in scattering theory (since
those can be incorporated "by hand") but rather the inexorable presence of
interaction-induced infinite vacuum polarization clouds in QFT.

Needless to add such a scheme is purely phenomenological since the
interactions are not given in terms of coupling constants but rather coupling
functions (interaction potentials) \cite{Co-Po}. An S-matrix with all the
above properties fulfills the requirements of a conjecture by Weinberg
\cite{W} although it does not lead to a QFT. If one adds the crossing property
(which has however no implmentation in a DPI setting), one can prove the
uniqueness of the inverse scattering problem, but the existence of a QFT
remains open \cite{unique}.

\textit{QFT and DPI are the only known settings in which a unitary,
Poincar\'{e} invariant and macro-causal can be derived} and which also have
been reasonably well understood from a conceptual/mathematical viewpoint. For
DPI the mathematical existence of models and their construction is handled in
terms of well-known functional analysis concepts as in ordinary QM. In case of
QFT this is much more difficult in view of the fact that the perturbative
series is divergent and the sometimes provable Borel resummability does not by
itself establish the existence. Therefore it is encouraging that its most
intrinsic (field coordinatization-independent) formulation in terms of
spacetime localized operator algebras has led to a nonperturbative existence
proof for a special family of interacting two-dimensional factorizable models.
Hopefully this will be the beginning of a new nonperturbative understanding
which at the end could be a realization of the old dream of an intrinsic
construction of models of QFT without the quantization "crutches".

The main aim of this article is to put forward arguments showing that string
theory is not what most people think it is, namely a theory of an infinite
collection of particles whose mass/spin spectrum originates from a string
which vibrates in spacetime. The idea that it generalizes the pointlike
localized fields of QFT is a metaphor based on the analogy with QM where such
a spectrum is associated with a vibrating string. But the intrinsic
localization concept in QFT which is different from the non-covariant
Born-Newton-Wigner localization forbids such an interpretation. Since
localization is a notorious difficult issue which led to many
misunderstandings, the discussion of localization of the objects of string
theory requires careful preparation. This will be the main theme of the next section.

\section{On-shell crossing and thermal properties from causal localization}

In order to attain a solid vantage point for a critique of string theory, it
is helpful to recall the issue of localization which constitutes the basis for
the formulation and \textit{interpretation} of local quantum physics. The
easiest access with the least amount of previous knowledge is through the
Wigner one-particle theory. Wigner discovered \cite{Wig} that irreducible
positive energy ray representations of the Poincar\'{e} group come in three
families: massive particles with half-integer spin, zero mass halfinteger
helicity representations and zero mass "infinite spin" representations. For
brevity we will sometimes refer to these families using numbers 1,2,3. Whereas
the first and the third family are rather large because their Casimir
invariants have a continuous range \footnote{Whereas for the massive family
this is the value of the mass operator, the continuous value in case of the
infinite spin family is the Casimir eigenvalue of the faithfully represented
Euclidean group $E(2)$ (the little group of a lightlike vector).}, the finite
helicity family has a countable cardinality labeled by the halfinteger
helicities. All up to present observed particles are in the first two
families. The fact that no objects have been observed which fit into the third
family should not mislead us into prematurely dismissing these positive energy
representations. Their physical properties are somewhat unusual and the
presence of apparently strange astrophysical dark matter of largely unknown
properties \cite{invisible} advise caution against prematurely dismissing such
representations. In the present context these objects mainly serve the purpose
to explain what \textit{indecomposable string-like localization} means.

The three families have quite different causal localization properties. Let us
first look at the one with the best (sharpest) localization which is the
representation family of massive particles. For pedagogical simplicity take
the Wigner representation of a scalar particle with the representation space
\begin{align}
&  H_{Wig}=\left\{  \psi(p)|\int\left\vert \psi(p)\right\vert ^{2}\frac
{d^{3}p}{2p^{0}}<\infty\right\} \\
&  \left(  \mathfrak{u}_{Wig}(a,\Lambda)\psi\right)  (p)=e^{ipa}\psi
(\Lambda^{-1}p)\nonumber
\end{align}
Now define a subspace which, as we will see later on, consists of wave
function localized in a wedge. Take the standard $t-x$ wedge $W_{0}%
=(x>\left\vert t\right\vert ,~x,y$ arbitrary) and use the $t-x$ Lorentz boost
$\Lambda_{x-t}(\chi)\equiv\Lambda_{W_{0}}(\chi)$%
\begin{equation}
\Lambda_{W_{0}}(\chi):\left(
\begin{array}
[c]{c}%
t\\
z
\end{array}
\right)  \rightarrow\left(
\begin{array}
[c]{cc}%
\cosh\chi & -\sinh\chi\\
-\sinh\chi & \cosh\chi
\end{array}
\right)  \left(
\begin{array}
[c]{c}%
t\\
z
\end{array}
\right)
\end{equation}
which acts on $H_{Wig}$ as a unitary group of operators $\mathfrak{u}%
(\chi)\equiv$ $\mathfrak{u}(0,\Lambda_{z-t}(\chi))$ and the $x$-$t$ reflection
$j:$ $(x,t)\rightarrow(-x$,$-t)$ which, since it involves time reflection, is
implemented on Wigner wave functions by an anti-unitary operator
$\mathfrak{u}(j).$ One then forms the unbounded\footnote{The unboundedness of
the $\mathfrak{s}$ involution is of crucial importance in the encoding of
geometry into domain properties.} \textquotedblleft analytic
continuation\textquotedblright\ in the rapidity $U_{Wig}(\chi\rightarrow
-i\pi\chi)$ which leads to unbounded positive operators. Using a notation
which harmonizes with that of the modular theory in mathematics \cite{Su}, we
define the following operators in $H_{Wig}$
\begin{align}
&  \delta^{it}=U_{Wig}(\chi=-2\pi t)\equiv e^{-2\pi iK}\label{pol}\\
\mathfrak{\ }  &  \mathfrak{s}=\mathfrak{\ \mathfrak{j}}\delta^{\frac{1}{2}%
},\mathfrak{\mathfrak{j}}=U_{Wig}(j),~\delta=\delta^{it}|_{t=-i}\nonumber\\
&  ~\left(  \mathfrak{s}\psi\right)  (p)=\psi(-p)^{\ast}\nonumber
\end{align}
Since the anti-unitary operator $\mathfrak{j}$ is bounded, the domain of
$\mathfrak{s}$ consists of all vectors which are in the domain of
$\delta^{\frac{1}{2}}.$ With other words the domain is completely determined
in terms of Wigner representation theory of the connected part of the
Poincar\'{e} group. In order to highlight the relation between the geometry of
the Poincar\'{e} group and the causal notion of localization, it is helpful to
introduce the real subspace of $H_{Wig}$ (the closure refers to closure with
real scalar coefficients).%

\begin{align}
\mathfrak{K}  &  =\overline{\left\{  \psi|~\mathfrak{s}\psi=\psi\right\}
}\label{K}\\
dom\mathfrak{s~}\mathfrak{=K}  &  +i\mathfrak{K},~\overline{\mathfrak{K}%
+i\mathfrak{K}}=H_{Wig},\mathfrak{K}\cap i\mathfrak{K}=0\nonumber
\end{align}

The reader who is not familiar with modular theory will notice that these
modular concepts are somewhat unusual and very specific for the important
physical concept of causal localization; despite their physical significance
they have not entered the particle physics literature. One usually thinks that
an \textit{unbounded} anti-linear involutive ($\mathfrak{s}^{2}=1$ on
$dom\mathfrak{s}$) operator which has two real eigenspace associated to the
eigenvalues $\pm1$ is an impossibilty, but its ample existence is the essence
of causal localization in QFT. Their conspicuous absence in the mathematical
physics literature is in a surprising contrast with their pivotal importance
for an intrinsic understanding of localization in local quantum physics (LQP).

The second line (\ref{K}) defines a property of an abstract real subspace
which is called \textit{standardness} and the existence of such a subspace is
synonymous with the existence of an abstract $\mathfrak{s~}$operator. This
property is the one-particle version of the Reeh-Schlieder property of QFT
\cite{Ha} which is also sometimes referred to (not entirely correct) as the
"state-field relation".

The important analytic characterization of modular wedge localization in the
sense of pertaining to the dense subspace $dom\mathfrak{s}$ is the strip
analyticity of the wave function in the momentum space rapidity $p=m(ch\chi
,p_{\perp},sh\chi).$ The requirement that such a wave function must be in the
domain of the positive operator $\delta^{\frac{1}{2}}$ is equivalent to its
analyticity in the strip $0<\chi<i\pi,$ and the action of $\mathfrak{s}$
(\ref{pol}) relates the particle wave function on the lower boundary of the
strip which is associated to the antiparticle wave function on the negative
mass shell.

This relation of particle to antiparticle wave functions is the conceptual
germ from which the most fundamental properties of QFT, as crossing, existence
of antiparticles, TCP theorem, spin-statistics connection and the thermal
manifestation of localization originate. Apart from special cases this fully
quantum localization concept cannot be reduced to support properties of
classical test functions.

More precisely the modular localization structure of the Wigner representation
theory "magically" preempts these properties of a full QFT already within the
one-particle sector; to be more specific: these one-particle properties imply
the corresponding QFT properties via time-dependent scattering theory
\cite{Mu}.

Hence a modification of of those fundamental properties, as the replacement of
crossing by Veneziano duality, is changing the principles of local quantum
physics (i.e. the result of more than half a century of successful particle
physics research) in a conceptually unsecured way. It was the first step into
the 40 year reign of metaphors in particle physics which culminated in the
wastful TOE.

The mentioned one-particle thermal aspects follows directly from (\ref{pol})
by converting the dense set $dom\mathfrak{s}$ via the graph norm of
$\mathfrak{s}$ into an Hilbert space in its own right $H_{G}\subset H_{Wig}$%
\begin{align}
\left\langle \psi\left\vert 1+\delta\right\vert \psi\right\rangle  &
=\left\langle \psi|\psi\right\rangle _{G}\label{an}\\
\left\langle \psi|\psi\right\rangle |_{dom\mathfrak{s}}  &  =\int\frac{d^{3}%
p}{2p_{0}}\frac{1}{1+e^{2\pi K}}\left\vert \psi_{G}(p)\right\vert ^{2}%
,~\psi_{G}\in H_{G}\nonumber
\end{align}
This formula represent the restriction of the norm to the strip analytic
function in terms of Hilbert space vectors $\psi_{G}$ which are free of
analytic restrictions. The result is the formula for a one point expectation
value in a thermal KMS state with respect to the Lorentz boost Hamiltonian K
at temperature $2\pi$. As we will see in a moment, the modular relation
(\ref{pol})\ in the Wigner one-particle setting is the pre-stage for the
crossing relation as well as an associated KMS property in an interacting
QFT\footnote{The thermal manifestation of localization is the strongest
seperation between QM and QFT \cite{interface}.}.

Before we get to that point, we first need to generalize the above derivation
to all positive energy representations and then explain how to get to the
sub-wedge (spacelike cones, double cones) modular localization spaces. For the
generalization to all positive energy representations we refer the reader to
\cite{B-G-L}\cite{MSY}; but since the sharpening of localization is very
important for our critique of string theory in the next section, it is helpful
to recall some of the points in those papers.

In the first step one constructs the "net" of wedge-localized real subspaces
$\left\{  \mathfrak{K}_{W}\right\}  _{W\in\mathcal{W}}.$ This follows from
covariance applied to the reference space$\mathfrak{K}_{W_{0}}.$ In the second
step one aims at the definition of nets associated with tighter localization
regions via the formation of spatial intersections
\begin{equation}
\mathfrak{K}(\mathcal{O})\equiv\cap_{W\supset\mathcal{O}}\mathfrak{K}_{W}
\label{inter}%
\end{equation}
Note that the causally complete nature of the region is preserved under these
intersections in accordance with the causal propagation principle which
attributes physical significance to the causal closures of regions (this is
the reason for the appearance of noncompact or compact conic regions in local
quantum physics). In this way localization properties have been defined in an
intrinsic way i.e. separate from support properties of classical test functions.

The crucial question is how "tight" can one localize without running into the
triviality property $\mathfrak{K}(\mathcal{O})=0.$ The answer is quite
surprising: For all positive energy representations one can go down from
wedges to spacelike cones $\mathcal{O=C~}$of arbitrary narrow size
\cite{B-G-L}
\begin{align}
&  \mathfrak{K}(\mathcal{C})~is\text{ }standard\\
&  \mathcal{C}=\left\{  x+\lambda\mathcal{D}\right\}  _{\lambda>0}\nonumber
\end{align}
i.e. the non-compact spacelike cones result by adding a family of compact
double cones with apex $x$ which arise from a spacelike double cone
$\mathcal{D}$ which touches the origin. Since there are three families of
positive energy Wigner representation\footnote{In d=1+2 there are also
plektonic/anyonic representations which will not be considered here.} one can
ask this question individually for each family.

The family with the most perfect localizability property is the massive one,
because in that case each $\mathfrak{K}(\mathcal{D})$ for arbitrary small
double cones is standard. On the opposite side is the third (massless infinite
spin) family for which the localization in arbitrarily thin spacelike cones
(in the limit semiinfinite strings) cannot be improved \cite{Yn}. The second
family (massless finite helicity) is in the middle in the sense that the
$\mathfrak{K}(\mathcal{D})$ spaces are standard but that the useful
"potentials" (vector potential in case of s=1,symmetric tensors for s=2) are
only objects in Wigner representation space if one permits spacelike cone
localized objects i.e. they covariant vectorpotentials cannot be associated
with compact spacetime regions.

In fact there exists a completely intrinsic argument on the level of subspaces
associated with field strengths which attributes a representation theoretical
property to these "stringlike" potentials. It turns out that\ "duality"
relation (Haag duality)
\begin{equation}
\mathfrak{K}(\mathcal{O})=\mathfrak{K}(\mathcal{O}^{\prime})^{\prime}%
\end{equation}
in massive representations holds for all spacetime regions including
non-simply connected regions. Here the dash on $\mathcal{O}$ denotes the
causal disjoint, whereas $\mathfrak{K}(\mathcal{O})^{\prime}$ is the
symplectic complement of $\mathfrak{K}(\mathcal{O})$ in the sense of the
symplectic form defined by the imaginary part\footnote{For halfinteger spin
there is a slight change.} of the inner product in $H_{Wig}$ This ceases to be
the case for zero mass finite helicity representation where there is a
\textit{duality defect} when \ $\mathcal{O}$ is multiply connected (example:
the causal completion of the inside of a torus at t=0)$.$ In that case one
finds
\begin{equation}
\mathfrak{K}(\mathcal{O})\subsetneqq\mathfrak{K}(\mathcal{O}^{\prime}%
)^{\prime}%
\end{equation}
which can be shown to be related to the string-like localization of potentials
\cite{MSY} i.e. this "defect" is the intrinsic indicator of the presence of
stringlike potentials.

These properties of localized Wigner subspaces can easily be converted to the
corresponding properties of a system (net) of spacetime indexed subalgebras of
the Weyl algebra or (for halfinteger spin) the CAR algebra. Since the reaction
between subspaces and subalgebras is functorial, all spatial properties have
their operator algebraic counterpart and one obtains (for simplicity we
restrict to the bosonic case)
\begin{align}
\mathcal{A(O})  &  \equiv alg\left\{  e^{i(a^{\ast}(\psi)+h.c.)}|~\psi
\in\mathfrak{K}(\mathcal{O})\subset H_{Wig}\right\} \label{mod}\\
SA\left\vert 0\right\rangle  &  =A^{\ast}\left\vert 0\right\rangle
,~A\in\mathcal{A(O}),~S=J\Delta^{\frac{1}{2}}\nonumber\\
\Delta^{it}\mathcal{A(O})\Delta^{-it}  &  =\mathcal{A(O}),~J\mathcal{A(O}%
)J=\mathcal{A(O})^{\prime}=\mathcal{A(O}^{\prime})\nonumber
\end{align}
where the operator-algebraic modular objects are the functorial images of the
spatial ones $S=\Gamma(\mathfrak{s}),~\Delta=\Gamma(\delta),~J=\Gamma
(\mathfrak{j}).$ It is important to not to misread the Weyl algebra generator
in the first line as an exponential of a smeared field; it is rather an
exponential of a (momentum space) Wigner creation/annihilation operator
integrated with Wigner wave functions from $\mathfrak{K}(\mathcal{O})$ i.e.
the functor uses directly the modular localization in Wigner space and does
not rely on the knowledge of pointlike quantum fields. Rather it represents an
intrinsic functorial construction of local algebras whereas the infinite
family of singular covariant fields result from the "coordinatization" of this
local net of algebras. The antiunitary involution $J$ not only maps the
algebra in its commutant (a general property of the T-T modular theory) but,
as a result of Haag duality, also brings the causal commutativity into the
game. Modular theory in the general operator algebra setting leads to the
action of the modular group $Ad\Delta^{it}$ which leaves the algebra invariant
and the action of the antiunitary involution $AdJ$ which transforms the
algebra into its Hilbert space commutant; both operators result from the polar
decomposition of the so-called (unbounded) Tomita involution $S$. \ The field
generators of this net of algebras are of course the well-known singular
covariant free fields whose systematic group theoretical construction directly
from the Wigner representation theory (except the massless infinite spin
representations) can be looked up in the first volume of \cite{Wei}. Far from
being a property which can be easily generalized of disposed of causal
localization is an inexorable aspect of a profound mathematical theory whose
consequential application leads to the recognition that inner- and spacetime-
symmetries are consequences of modular positioning of algebras \cite{Ha}%
\cite{interface}.

For a bona \ fide string-localization which contrasts the string theory
metaphor, the third Wigner representation family is particularly useful. The
history of its unravelling is an interesting illustration of the intricacies
of localization \cite{interface}. This class of representations have
\textit{no pointlike generators}; in fact their compactly localized subspaces
are trivial $\mathfrak{K}(\mathcal{O})=0$ whereas the spacelike cone localized
subspaces $\mathfrak{K}(\mathcal{C})$ are standard i.e. $\overline
{\mathfrak{K}(\mathcal{C})+i\mathfrak{K}(\mathcal{C})}=H_{Wig}.$ the finite
helicity representation for which onlycertain tensor fields (for s=1 the
vectorpotentials) are stringlike localized, the QFT of the infinite spin
representation have no pointlike generators at all and there are strong
indication that there are also no compactly localized subobjects.

To make contact with the standard field formalism one looks at the
(necessarily singular) generators of these algebras. For the first two
families these are pointlike covariant fields $\Psi(x)$ apart from the finite
helicity potentials which, similar to the generators of the infinite spin
class, are described by string-localized field generators $\Psi(x,e)$ (leaving
off the tensor/spinorial indices) which depend in addition to a point x in
d-dimensional Minkowski spacetime also on a point in a d-1 dimensional de
Sitter space (the spacelike string direction) $e.$ The stringlike localization
nature shows up in the support properties of the commutator for whose
vanishing it is not sufficient that the starting point $x$ and $x^{\prime}$
are spacelike but rather
\begin{equation}
\left[  \Psi(x,e),\Psi(x^{\prime},e^{\prime})\right]  =0\text{~}%
only\text{~}for~x+\mathbb{R}_{+}e~><~x^{\prime}+\mathbb{R}e^{\prime}
\label{string}%
\end{equation}
This string-localization is real, unlike that in string theory which is
stringlike in a metaphoric but not in \ material spacetime sense (section 6).

The basic difference between the second (finite helicity) and third kind
Wigner representation type is that the string localization is only required in
relations in which the vectorpotentials play an important role whereas in case
of the third kind one does not expect the presence of pointlike localized composite.

The theory also says that there is no need to introduce generators which have
a higher dimensional localization beyond point- or semiinfinite string-like.
Note that it is of course not forbidden to introduce decomposable string (and
higher) localized operators as e.g.
\begin{equation}
\int\Psi(x)f(x)d^{4}x,\text{ }suppf\subset tube
\end{equation}
in the limit when the thickness of the tube approaches zero. When we talk
about semiinfinite string localization without further specification we mean
indecomposable strings. These are strings which in contrast to decomposable
strings cannot be observed in a counter since any registration device would
inevitably partition the string into the part inside and outside the counter
which contradicts its indecomposable nature (this is of course a metaphorical
argument which is in urgent need of a more explicit and intrinsic
presentation). The string-localized generators of the Wigner infinite spin
representation do not even admit pointlike localized composites i.e. net of
spacelike cone localized algebras has no compactly localized nontrivial
subalgebras. \ A milder form of string-like generation of representations
occurs for the zero mass finite helicity representation family which in some
way behaves localization-wise as standing in the middle between massive
representation (which are purely point-localized) and the third kind. These
representations are fully described in terms of pointlike localized field
strength but already before using these representations in interactions it
turns out that the additional introduction of "potentials" is helpful. Whereas
in the interaction-free case there is a linear relation between the observable
field strength and its potential whose inversion permits to rewrite the latter
as one or more line integral over the former, this feature is lost under
suitable interactions i.e. the string localized potential may become an
indecomposable string localized generator which cannot be approximated by
compactly observables. In this case the "visibility" of such objects in
counters with finite localizability becomes an issue.

In the presence of interactions there is no \textit{direct} algebraic access
to problems of localization from the Wigner one-particle theory. In the
Wightman setting based on correlation functions of pointlike covariant fields,
the modular theory for the wedge region has been derived a long time ago by
Bisognano and Wichmann \cite{Bi-Wi} and more recently within the more general
algebraic setting by Mund\footnote{That derivation actually uses the modular
properties of the Wigner setting which is connected via scattering theory to
the interacting wedge-localized algebras and then as explained above (via
intersection) to the modular structure of all local algebras $\mathcal{A(O}%
).$} \cite{Mu}. The resulting modular S-operator has the same property as in
(\ref{mod}) i.e. the "radial " part of the polar decomposition of the modular
involution $S$ is determined solely by the representation theory of the
Poincar\'{e} group i.e. the particle content whereas the $J$ turns out to
depend on the interaction \cite{Ann} since it is related to the scattering
operator $S_{scat}$%
\begin{equation}
J=J_{0}S_{scat} \label{J}%
\end{equation}
which in this way becomes a relative modular invariant between the interacting
and the free wedge algebra\footnote{$J_{0}$ is (apart from a $\pi$-rotation
around the z-axis of the t-z wedge) the TCP operator of a free theory and $J$
is the same object in the presence of an interaction.}. There is no change in
the construction of the $\mathcal{A(O})$ by intersecting $\mathcal{A(}W)s.$
However in the presence of interactions the functorial relation between the
Wigner theory gets lost. In fact no subwedge-localized algebra contains any
associated PFG (polarization-free-generator) i.e. an operator which creates a
pure one particle state from the vacuum (without an additional vacuum
polarization cloud consisting of infinitely many particle-antiparticle pairs).

Since the crossing property played a crucial role in S-matrix approaches to
particle physics, it is instructive to spend some time for its appropriate
formulation and on its conceptual content. Its most general formulation is
given in terms of formfactors which are products of W-localized operators
$A_{i}\in\mathcal{A}(W)$\footnote{Since all compactly localized operators can
be translated into a common W and since the spacetime translation acts on in
and out states in a completel known way this is hardly any genuine
restriction.} between incoming ket and outgoing bra states
\begin{align}
&  ^{out}\left\langle p_{k+1},p_{k+2},...p_{n}\left\vert A\right\vert
p_{1},p_{2},..p_{k}\right\rangle ^{in}=\label{cross}\\
&  ^{out}\left\langle -\bar{p}_{k},p_{k+1},p_{k+2},...p_{n}\left\vert
A\right\vert p_{1},p_{2},..p_{k-1}\right\rangle ^{in},~A=\Pi_{l}A_{l}\nonumber
\end{align}
where the crossed particle is an outgoing anti particle relative to the
original incoming particle Hence all formfactors of $A$ with the same total
particle number n are related to one \ "masterfunction" by analytic
continuation through the complex mass shell from the physical forward shell to
the unphysical backward part. Hence the predictive power of crossing is
inexorably connected with the concept of analytic continuation i.e. it is
primarily of a structural-conceptual kind. It is convenient to take as the
master reference formfactor the vacuum polarization components of $A\Omega$
i.e. the infinite system of components of the infinite vacuum polarization
cloud of $A\Omega.$ Needless to add that the crossing relation may be empty in
case that the operator $A~$cannot absorb the energy momentum difference
between the original value and its continued negative backward mass shell
value. In this setting the S-matrix arises as a special case for
$A=\mathbf{1}$ i.e. an operator which cannot absorb any nontrivial energy
momentum. In this case it is not possible to use the vacuum polarization as a
reference and neither leads the crossing of one momentum in the 2-particle
elastic amplitude to a meaningful relation (but the simultaneous crossing of
two particles in the in and out configuration is meaningful).

This is also the right place to correct the picture of the QFT vacuum as a
bubbling soup which for short times, thanks to the Heisenberg uncertainty
relation between time and energy, can violate the energy momentum
conservation\footnote{The origin of these metaphors sees to be the too literal
interpretation of the momentum space Feynman rules.}. The correct picture is
that (modular) localization in QFT costs energy-momentum i.e. in order to
split the vacuum into a product vacuum%

\begin{align}
&  \omega(A_{1}A_{2})\rightarrow\omega(A_{1})\omega(A_{2}),~A_{i}%
\in\mathcal{A}(\mathcal{O}_{i})\\
&  \Omega\rightarrow\Omega\otimes\Omega,~~A_{1}A_{2}\rightarrow A_{1}\otimes
A_{2}\nonumber
\end{align}
where the index 1 refers to a spacetime region $\mathcal{O}_{1}$ and 2 labels
a region $\mathcal{O}_{2}$ in the causal complement i.e. $\mathcal{O}%
_{2}\subset\mathcal{O}_{1}^{\prime}.$Whereas the tensor factorization in QM
for disjoint regions at a fixed time is automatic, the tensor product vacuum
in QFT is a highly energetic thermal state whose energy diverges in the limit
when the closure of $\mathcal{O}_{2}$ touches $\mathcal{O}_{1}^{\prime}.$ The
difference with respect to tensor factorization can be traced back to the
conceptual difference between Born- and modular- localization. The tensor
factorization in QM leads to an information theoretical entanglement whereas
the entanglement for the QFT tensor factorization is related to thermal
properties \cite{interface}. Confusing the two is the main cause for the
"information paradox". \ 

The image of the bubbling vacuum shows that metaphors are not limited to
string theory. But those used in QFT do usually not lead to serious misunderstandings.

The origin of the formfactor crossing property (\ref{cross}) lies in the strip
analyticity of wedge localized states and correlation function. For wedge
localized wave functions this was explained above (\ref{pol}, \ref{an}). For
simplicity let us limit the interacting situation to the simplest case
\begin{align}
\left\langle 0\left\vert A\right\vert p\right\rangle  &  =\left\langle
-\bar{p}\left\vert A\right\vert 0\right\rangle \\
\left\langle 0\left\vert AB\right\vert 0\right\rangle  &  =\left\langle
0\left\vert B\Delta A\right\vert 0\right\rangle \nonumber
\end{align}
where in the second line we have written the KMS property for the wedge
algebra which is a general consequence of modular operator theory and for the
special case of wedge localization agrees with Unruh's observations about
thermal aspects of Rindler localization ($\Delta^{it}=U_{W}^{boost}(\chi=-2\pi
t)$). But how to view the first relation as a consequence of the second? The
secrete is that although the intersection of the space of one-particle states
with that obtained from applying compact localized algebras to the vacuum (and
closing in the modular graph norm) is trivial, that with the noncompact
wedge-localized algebra is not; it is even dense in the one particle Hilbert
space. Once it is understood that there exists a wedge affiliated operator $B$
which, if applied to the vacuum, generates the one-particle state, one can
apply the KMS relation in the second line. The rest the follows from
transporting the left side $B$ as $B^{\ast}$ to the bra vacuum. The rest
follows by rewriting the $B^{\ast}\left\vert 0\right\rangle $ as $SB\left\vert
0\right\rangle $ using modular operator theory and using (\ref{J}). The
resulting $\Delta^{\frac{1}{2}}JB\left\vert 0\right\rangle =\Delta^{\frac
{1}{2}}J_{0}B\left\vert 0\right\rangle $ (since the $S_{scat}$ matrix acts
trivially on one-particle states) leads to the desired result\footnote{The
plane wave relation should be understood in the sence of wave packets from the
dense set of strip-analytic wave functions.} $\Delta^{\frac{1}{2}}%
J_{0}\left\vert p\right\rangle =\left\vert -p\right\rangle .$ The general form
(\ref{cross}) would follow if we could generalize the KMS relation to include
operators from the wedge localized in and out free field algebras. They share
with $\mathcal{A}(W)$ the same unitary Lorentz boost as the modular group but
their modular inversions $J$ are not equal and hence additional arguments are
required. We will leave the completion of the derivation of crossing to a
future publication \cite{M-S}.

To criticize the string theory interpretation of the canonically quantized
Nambu-Goto mode we do not have to go into subtle details. For such bilinear
Lagrangians (leading to linear Euler-Lagrange equations) the connection
between localization of states and locality of operators is that in free field
theory. In this case it is possible to pass from the "first quantized" version
directly to its "second quantization" i.e. to the N-G "string field theory".
Since the physical content consists of an infinite tower of massive particles
(with one layer of finite helicity massless representations), the only
question is does the original classical parametrization lead to fields which
are decomposable strings or are they point localized? In the first case one
could resolve the composite string in terms of a "stringy spread" of
underlying pointlike fields\footnote{A (composite) string from a Lagrangian
setting would be surprising.} whereas in the second case the string
terminology would only refer to the classical origin and lack any intrinsic
quantum meaning. The suspense will be left to the next section.

\section{A turn with grave consequences}

Although the partisans of the S-matrix bootstrap program placed the new and
important crossing property into the center of their S-matrix setting, they
failed to come up with a constructive proposal which could implement this new
requirement. Other older requirements, as Stueckelberg's macro causality, were
not mentioned in the bootstrap program, they were probably forgotten in the
maelstrom of time. The important question in what way (on-shell) crossing is
related to the causality principles of QFT did not receive the attention it
merits; it has no place in an ideology which set out to cleanse particle
physics from the dominance of QFT. In fact most of the efforts were focussed
on the elastic scattering amplitude on which Mandelstam's conjecture
(concerning the validity of a certain double spectrum representation) was the
central object in terms of which the crossing had a simple natural formulation.

The crossing property was verified in the Feynman perturbation theory where a
certain analytic continuation from momenta on the forward to the backward
shell changes the in/out association of the external legs of Feynman graphs
within the same perturbative order. As a result all physical channels with the
same total number of external lines are just different boundary values of one
analytic master function. Since the analytic properties of Feynman graphs are
established without much efforts, it appears at first sight that the
perturbative version of crossing is an easy matter. However the condition to
perform this analytic continuation inside the complex mass shell adds some
nontrivial aspects. By looking at the S-matrix of 2-dimensional factorizing
models one can see that crossing involves a delicate interplay between
one-particle poles and the multi-particle scattering continuum \cite{Ba-Ka}.

From the motivation in the original papers it is quite evident that Veneziano
\cite{Venez} had this kind of crossing in mind when he set out to construct an
explicit implementation within the Mandelstam setting for 2-2 elastic
scattering amplitudes. But being guided by the properties of $\Gamma
$-functions he arrived at a crossing in terms of an infinite set of particles
forming a mass/spin tower. This formal version of crossing in which the higher
particle scattering states (multiparticle "cuts") is not consistent with the
principles underlying QFT. The "Veneziano duality" marked the beginning of a
new setting in particle physics which was defined in terms of metaphoric
recipes but for which one lacked an understanding about the intrinsic physical
meaning \footnote{The realization of duality with the help of identities
between Gamma functions is the only successful "by hand" construction within a
pure S-matrix scheme. Ironically it did not lead to a model for the crossing
property, but it created a new concept of "duality" which is not supported by
any established principle in particle physics and marks the begining of the
metaphoric ascent into string theory.}. This point will be elaborated in the
context of its string theoretic formulation in the next section.

At the time of its discovery the difference between duality and crossing did
not cause much headache since the idea of an infinite mass/spin spectrum
needed in order to implement duality was favored from a phenomenological point
of view. There is of course nothing unusual to explore all phenomeological
consequences and leave the more conceptual problems to times after the
phenomenological success has been secured beyond any doubt. The Regge-pole
dominance gained a lot of popularity before it was ruled out by experiments
which favored the quantum field theory of QCD as the appropriate description
of strong interactions.

Veneziano's remarkable observation about the existence of a rather simple
mathematically attractive idea, which is capable to generate an interesting
mass/spin spectrum from a suitably formulated duality requirement, continued
to attract attention even after the Regge pole idea lost its attraction.
Although not noticed at the time, the quest for a field equation for an
infinite component wave function\footnote{The motivation originated from the
successful algebraization of the hydrogen spectrum in terms of
represesentation of the noncompact $O(4,2)$ group. In the relativistic case
the group theoretical Ansatz did not lead to interesting results. String
theory replaces group theory by an infinite collection of oscillators which
maintain the pointlike localization (section 6)..} with an infinite mass/spin
content, which remained an unfulfilled dream in previous attempts at infinite
component relativistic field equations, found a successful implementation in
Veneziano's dual model and in the subsequent string theory..The Veneziano
Ansatz did not only create a mass/spin tower. It also led to explicit analytic
expressions for amplitudes which, leaving aside problems of unitarity, had the
formal appearance of an approximation to the elastic part of a "could be" S-matrix.

Extending the search for an implementation of duality-based on properties of
Gamma function, Virasoro \cite{Vir} arrived at a model with a different and
somewhat more realistic looking particle content. The duality setting became
more complete and acquired some additional mathematical charm after it was
extended to n particles \cite{DHS}. The resulting "dual resonance model" was
the missing link from the phenomenological use of Gamma function properties to
a conceptually and mathematically attractive formulation in terms of known
concepts in chiral conformal QFT, the new idea being that Minkowski spacetime
should be envisaged as the "target space"\footnote{Note that the notion of
target space is well defined only in classical field theories (where fields
have numerical values) whereas in QFT its meaning is metaphorical.} of a
suitably defined chiral model.

It is worthwhile to look at the mathematical formulation and the associated
concepts in some detail. The conformal model which fits the dual resonance
model are the charge creating fields of a multi-component abelian chiral
current which are customarily described in the setting of bosonization
\begin{equation}
\Psi(z,p)=:e^{ip\phi(z)}:
\end{equation}
where the d-component $\phi(z)$ is the formal potential of a d-component
chiral current $j(z)=\frac{d}{dz}\phi(z),$ and p is a d-component numerical
vector whose components describe (up to a shared factor) the value of the
charge which the $\Psi$ transfers$.$ The fact that the Hilbert space for
$\Psi$ is larger than that of the current $j$ is the place where the language
of bosonization becomes somewhat metaphoric; this point is taken care of by a
proper quantum mechanical treatment of zero modes which appear in the Fourier
decomposition of $\phi(z).$ The n-point functions of the $\Psi$ define the
integrands of the scattering amplitudes of the dual resonance model; the
latter result from the former by z-integration after multiplication with
$z_{i}$ dependent factors \cite{Vec}. \ Hence the energy-momentum conservation
of the target theory results from the charge conservation of the conformal
source theory. Even the cluster property is preempted on the level of the
conformal current model. The verification of the causal one-particle structure
is more tricky, but it also can be traced back to a property of the charge
composition structure, but it is a very special property which has no direct
relevance within the logic of a current algebra.

The crucial additional step which secured the survival of the mathematical
formalism was the streamlining it received by connecting it to the
\textit{Lagrangian of a Nambu-Goto string}. It is this step which facilitated
the formulation of prescriptions for higher order interactions; but these
recipes are, unlike in the Lagrangian setting of QFT, not consequences of a
Lagrangian formulation . Since the prescription for interactions is on-shell
i.e. directly formulated in terms of the particle/spin tower and therefore
should be interpreted as a contribution to the S-matrix, it is difficult to
judge whether it is physically acceptable since physical properties for the
S-matrix are hard to formulate and even harder to check. One could of course
try to verify the macrocausality properties. This has not been done; but
assuming that it works, one can ask furthergoing question about unitarity.
There are conjectures how to unitarize amplitudes with the help of geometric
concepts involving Riemann surfaces of higher genus, but again there is no
tangible result which comes close to unitarity checks in the Feynman setting.
The same holds for the question whether the duality on the one-particle level
persists to higher orders.

One of its most curious aspects is that string theory only exists in the
positive energy setting in d=10 in the presence of supersymmetry. The origin
of this severe restriction is the identification of the physical spacetime
with the target dimension of a multicomponent chiral current model. It turns
out that this is only possible with 10 target dimensions. However even this
weird looking possibility represents a pyrrhic victory for the conformal
embedding construction: instead of a stringlike object one obtains an infinite
component pointlike generated wave functions. This wave function is not
string-localized since all the degrees of freedom of the chiral theory have
become "inner" i.e. live over one localization point.

Truely spacetime localized strings exist in any dimension $\geq$ 3, so what
does this distinction of 10 dimensions mean in physical terms? The resolution
of this paradoxical situation is that that the objects associated with the
embedding of chiral theories or from the quantization of the Nambu-Goto
Lagrangian are not strings in the spacetime sense as their protagonists
imagine and present them, rather they describe nontrivial solution of a
slightly older problem of constructing infinite component wave
functions/fields which requires that special spacetime dimension in order to
exist at all. The oscillators from the Fourier decomposition of a string
indeed play an important role in setting the mass/spin spectrum, but they have
nothing to do with a stringlike localization in spacetime but "live" in a
Hilbert space which is attached to a point in an analogous sense as the spin
components are attached to a pointlike localized field or wave function.

To the extend that one calls degrees of freedom which do not affect the
localization of objects of local quantum physics "inner", the harmonic
oscillator variables are indeed inner, in contrast to the variable $e$ which
characterises the string direction. It is however interesting to note that
these kind of infinite degree of freedom inner oscillator variables accomplish
the construction of a nontrivial 10-dimenional infinite component wave
function where previous attempts based on noncompact groups (in analogy to the
dynamical $O(4,2)~$group which describes the hydrogen spectrum) failed (in any
dimension). \ This raises the curious question why the infinite component
program fell out of favor, but not the metaphoric string theory progam, or to
phrase it in a more specific way: why is the formulation of interactions in a
metaphoric setting more acceptable than in an intrinsic formulation? A
somewhat ironies answer may be that it is only in the metaphoric setting the
graphical tube rules for the transition amplitudes have some intuitive appeal.
The proofs of the infinite componnt wave function content of the Nambu-Goto
model will be presented in the next section.

In the aforementioned conformal field theoretic description based on the use
of abelian currents the reason why most spacetime dimensions are excluded have
some plausibility. In that setting the spacetime symmetry is connected to the
inner symmetry of abelian charges. Inner symmetries are from experience
related to compact groups; for QFT of at least 3 spacetime dimensions this can
be rigorously established \cite{Ha}. On this basis one would conclude that
there is no possibility of having vector-valued conformal fields on which a
Lorentz transformation can act. This argument is for several reasons too
naive\footnote{One reason being that there is no such theorem in low
dimensional theories.}. But as the existence of the N-G model shows, it is
only possible under extremely special circumstances.

The dual resonance model and its extension to string theory\footnote{Modern
string theory has moved a far distance beyond the N-G model. In textbooks and
reviews this model only serves to illustrate the underlying idea as well as
its historical origin. Since our critical view is not directed towards
concrete calculations but aims at its metaphoric interpretation, it suffices
to exemplify our point in this model. The crucial question do results mean
which are derived on the incorrect metaphor of a string in spacetime.} was,
for the same reasons as the Regge pole model, soon contradicted by
experiments; despite all its curious and in some cases surprising properties
nature, as far as it is known, had no use for the scattering amplitudes coming
from string S-matrix prescriptions. Like the Regge pole-model it could have
ended in that dustbin of history reserved for curious but not useful observations.

The sociological situation changed drastically when, as a result of containing
all possible spins, some physicists had the audacity to propose superstring
theory as a theory of everything (TOE). By identifying the quantum aspects of
gravity with the zero mass s=2 component in the mass/spin tower and
emphasizing that it promises to lead to a totally convergent S-matrix
essentially without free parameter (this is apparently true in lowest
nontrivial order) it acquired a hegemonic status above QFT.

By its very nature an S-matrix cannot have ultraviolet divergencies, but the
claim that it is free of any additional parameters beyond the string tension
would be remarkable, always assuming that the tube geometry (which constitutes
the graphical visualization of interacting strings) admits an interpretation
which avoids the metaphoric reading of a string in spacetime (see next
section). What is surprising about this proposal is not that it was made; the
speculative dispositions of theoretical physicists lead to many fantastic
suggestions, some of them even enter journals. The surprising aspect is rather
that the string idea was embraced quite uncritically by leading theoreticians
so that the formation of a superstring community around this idea became
inevitable. The resulting sociological setting was very different from
previous situations in which the established members of the particle physics
community provided the crucial balancing criticism. In this way the march into
a theory which has a highly technical mathematical formalism and a totally
metaphoric physical interpretation took its course.

Modern string theory has moved a far distance beyond the N-G model. In
textbooks and reviews this model only serves to illustrate the underlying idea
as well as its historical origin. Since our critical view is not directed
towards concrete calculations but aims at their metaphoric interpretation, it
suffices to exemplify our point in this model. All post N-G advancements of
string theory, as sophisticated as they may be, are afflicted with one
original sin: \textit{the incorrect metaphor of strings in spacetime. }The
crucial question, namely what do results real mean which are derived on the
incorrect metaphor of a string in spacetime remained unanswered. The only way
in which one can rescue the string theoretic calculations is to find out what
remains if this metaphor is replaced by its true meaning. This is a problem
which will be analyzed in the concrete context of the N-G model in the next section.

\section{The ascent of the metaphoric approach to particle physics: string
theory}

When it became clear that QCD was the more appropriate theory for the
description of strong interactions, a completely new and physically more
relevant playground offered new challenges for particle theorists and
phenomenologists. As a result the era of string theory which started with the
dual model came to an abrupt but only temporary end. In this way string theory
became an "orphan" of particle physics; its modest mathematical charm, which
consisted in Euler type of identities between gamma functions, lost its
physical attraction.

If it were not for a small community of string theorists who remained unshaken
in their belief that hidden behind the many unexpected properties there ought
to exist a deep new kind of quantum physics, string theory, as we presently
know it, would not be around today. Although those few years when string
theory was out of the limelight would have been the right time to look at its
conceptual foundations, this is not what actually happened. Rather than a
foundational critical review, the main attraction was the idea that a theory,
which by its "stringy" nature contains a mass/spin tower in which all possible
spin values occur (including that of a spin=2 graviton), could lay claim to
represent a TOE i.e. a unique theory of all quantum matter in the universe.
Its apparent uniqueness in form of a superstring supported such a belief, at
least in the beginning. In this way the chance to remove from string theory
the metaphoric casing of its birth was lost and the delicate critical
distinction between the autonomous conceptual content of a theory and the
metaphoric presentation of its computational rules became increasingly
blurred. It is our aim in this section to elaborate on this point.

In the previous section it was pointed out that the duality property, which
led to string theory, came into being by mathematical expediency; as a result
its content has little in common with the crossing property in QFT which is a
consequence of physical principles. It is not uncommon in particle theory to
substitute a problem which one cannot solve by a similar looking one which is
more susceptible to solution. One gets into very unsafe waters if the
construction has a high degree of mathematical consistency within a
fundamentally flawed physical concept.

As will be seen in the sequel the idea that string theory has to do with
string-like extended objects in spacetime is fundamentally flawed even though
no geometric-mathematical property of strings is violated. Geometry does not
care about its physical realization\footnote{A similar but less extreme
situation is met with the mathematics of Riemann surfaces which occurs at
different places which have nothing to do with surfaces in a physical sense
e.g. Fuchsian groups. In chiral conformal theory one often finds the
expression "chiral theory on Riemann surfaces" when the analytic continuation
of the correlation functions yields a Riemann surface, but this should not be
confused with the physical living (localization) space which remains always
one-dimensional.}. By not noticing that the real autonomous content
contradicts the metaphor, the problem becomes compounded. The problem arises
from confusing geometric properties of string theory which are correct in
their own geometric-mathematical right, with physical-material localization in
spacetime. For precisely this reason the Atiyah-Witten geometrization of
particle physics which started at the end of the 70s was a double-edged sword.
Since the Lorentz-Einstein episode we know that the physical interpretation is
not an automatic consequence of mathematics.

In fact it was an uneasy feeling that string theory was constructed with an
excess of sophisticated "tinkering" and a lack of guiding principles which in
many theoreticians, especially those with a rich experience with conceptual
problems of QFT, nourished the suspicion that, even leaving aside the problem
of predictive power, there is something deeply surreal about this theory.

Instead of entering a point for point critique of the extensive and
technically laborious content of string theory, I will focus my critical
remarks to what I consider the Achilles heel of string theory, namely its
metaphoric relation with those localization concepts which are central for the
formulation and interpretation of particle physics.

We know from Wigner's representation theoretical classification (section 4)
that the indecomposable constituents of positive energy matter are coming in
three families: the massive family which is labeled by a continuous mass
parameter and a discrete spin, a discrete massless family with discrete
helicity, and finally a continuous zero mass family of with an infinite spin
(helicity) tower. Whereas theories involving the first two families have
generating pointlike localized fields or field strengths (with possibly
stringlike potentials), there are no pointlike covariant generators within the
last family; rather the sharpest localized generators in that case are
semiinfinite strings localized along the spacelike half-line $x+\mathbb{R}%
_{+}e,$ where $x$ is the starting point of the string and $e$ is the spacelike
direction in which it extends to spacelike infinity. Their localization shows
up in their commutation relation (\ref{string}).

Stringlike localized objects can of course also be constructed in pointlike
QFTs; one only has to spread a physical pointlike field along a string where
the spreading has to be done in the sense of distribution theory since the
resulting stringlike objects is still singular. Such a string will be referred
to as composite or \textit{decomposable; }it plays no role in structural
investigations\textit{.} Neither these strings nor their indecomposable
counterpart arise directly from Lagrangians, but their presence is important
for the understanding of the physical content of a theory. A excellent
illustration for an indecompsable string\footnote{Only by using unphysical
fields the DJM formula has the appearance of a decomposable string.} is
provided by the nonlocal aspects of electrical charge-carrying fields in QED
\cite{infra} whose sharpest localization is that of a Dirac-Jordan-Mandelstam
(DJM) semiinfinite spacelike string. There are also situations in which only
stringlike semiclassical solutions (whose quantum status remained unknown)
play a role.

Even in cases where the full physical content can be expressed in terms of
pointlike localized fields, there may be important physical reasons for their
introduction which are related to \textit{improvements in the short distance
behavior}, thus permitting an extension of the family of renormalizable
interactions. It turns out that covariant string-like generating free fields
exist for each all spins. Whereas the short distance dimension of point-like
fields increases with their spin, the short distance dimension (ssd) of their
string-like counterpart $\Phi(x,e)$ remains at $sdd\Phi=1$ \cite{MSY}. Hence
the formal power-counting criterion is fulfilled within the setting of
maximally quartic interaction$_{{}}$, and the remaining hard problem consists
in generalizing the rules of the perturbative Epstein-Glaser iteration from
point-like to string-like fields. This problem, which has obvious implication
for a radical reformulation of gauge theory\footnote{The present formulation
only contains locally gauge invarant observables within its formalism,
nonlocal gauge invariant objects have to be introduced "by hand"
\cite{infra}.} is presently being studied \cite{M-S}. \ 

As explained in section 4 (\ref{string}), the representations of the third
kind of positive energy matter (Wigner's famous infinite spin representations)
are \textit{indecomposable strings }\cite{MSY}; in fact these free string
fields, unlike those charge-carrying QED strings, do not even lead to
pointlike composites \cite{invisible}.These representations are therefore
excellent illustrations for the meaning of \textit{indecomposable stringlike
localized fields.}

The important lesson from section 4 is that \textit{localization is an
autonomous quantum theoretical concept} i.e. there is no general
correspondence to classical localization, although for pointlike localized
fields the Lagrangian quantization shows that a classical field at a point
remains also pointlike in the autonomous notion of quantum causal
localization. The latter comes with its own \textit{modular localization
formalism} which is of a representation theoretic kind and bears no trace of
any quantization parallelism (section 4). There is no negation of the fact
that the coincidence of the two was one of the luckiest moments in the history
of particle physics; QFT could and did start with Jordan and Dirac and did not
have to wait for the arrival of Wigner's representation theory.

The conceptual autonomy of quantum localization is underlined by the rather
involved arguments which are necessary to show the equivalence of the
quantization- with the representation-method \cite{Wei}. The representation
theoretical setting together with modular localization is more flexible, since
it immediately leads to infinitely many covariant field realizations for the
unique $(m,s)$ representation; from the quantization viewpoint this would be
hard to see. In fact the confusing situation during the 30s (when there were a
large number of proposals for field equation which looked different but
nevertheless were physically equivalent) was Wigner's motivation for taking
the unique representation theoretical route.

The correspondence breaks down in two cases which both turn out to be
string-localized. The more spectacular of the two is the mentioned infinite
spin representation where there is no relation at all to the classical spinor
formalism. The zero mass finite helicity case on the other hand is pointlike,
but not all spinorial objects which relate the physical spin with the formal
dotted/undotted spinorial; spin are admissible in the sense of
(\ref{admissible}); it is well-known that the physical photon representation
leads to field strength but not to a vectorpotential. Allowing string
localized fields one recovers all undotted/dotted spinorial possibilities
which were admissible in the massive case.

The long time it took to find a covariant field description for the infinite
spin representations finds its explanation in the nonexistence of any
pointlike generating field. More recently, with a better understanding of
modular localization (section 4), it became clear that the generating fields
with the best localization are covariant string-like fields \cite{MSY} which
generate indecomposable string states. It serves as an excellent illustration
of what string-localization in a non-metaphoric intrinsic sense really means.

This intrinsicness stands in an ironic constrast to the fact that the
classical relativistic Nambu-Goto Lagrangian does not lead to quantum object
which are string localized; its full content is rather described by a
point-like localized \textit{infinite component field}, a special case of a
generalized free field. The main topic for the rest of this section will be to
demonstrate that the meaning of "string" in connection of "string theory" is
purely metaphoric.

Metaphors in particle theory are often helpful as intermediate crutches
because they facilitate to find the correct concepts and their appropriate
geometric-mathematical implementation. It is not important that they contain
already the correct physical interpretations of the object which has been
constructed, it suffices that they helped to get there. After the object has
been constructed it unfolds its own intrinsic interpretation. QFT is an
excellent illustration; no matter if one gets to it either by quantization or
representation theory, it always unfolds its intrinsic logic which rejects any
outside metaphoric interpretation that does not agree with its autonomous
meaning. This internal strength it owes to its \textit{quantum causal
localization} (which only theories with a maximal velocity but not QM are able
to possess). If one gets into a situation where metaphors contradict the
intrinsic properties and one fails to notice this lack of balance, one runs
into a very serious problem.

A good starting point for the problem of localization in string theory is to
remind oneself of the form of the most general pointlike field which is free
in the sense of leading to a c-number (graded) commutator. In case of a
discrete mass spectrum, a "masterfield" which contains all spins as well as
all possibilities to interwine a given physical spin s with all admissible
covariant representations looks as follows
\begin{align}
&  \Psi(x)=\sum_{\left(  A\dot{B},s\right)  ,i}\Psi^{(A\dot{B},s)}%
(x,m_{\left(  A\dot{B},s\right)  ,i})\label{admissible}\\
&  \Psi^{(A\dot{B},s)}(x,m_{\left(  A\dot{B},s\right)  ,i})=\frac{1}{\left(
2\pi\right)  ^{\frac{3}{2}}}\int e^{-ipx}\sum_{s_{3}=-s}^{s}u^{\left(
A\dot{B},s\right)  ,i}(s;p,s_{3})\times\\
&  \times a(s;p;s_{3})\frac{d^{n-1}p}{2\sqrt{\vec{p}^{2}+m_{\left(  A\dot
{B},s\right)  ,i}^{2}}}+c.c.\nonumber\\
&  \left[  \Psi(x),\Psi^{\ast}(y)\right]  _{grad}=\sum_{s}\sum_{A,\dot{B}%
\in\left(  A\dot{B},s\right)  }\int\mathbf{\Delta}^{(A\dot{B},s)}%
(x-y;m_{\left(  A\dot{B},s\right)  ,i},s)
\end{align}
The meaning of the notation is as follows:

\begin{description}
\item The$\Psi_{\left(  A\dot{B},s\right)  }^{(A\dot{B})}(x,m_{\left(
A\dot{B},s\right)  ,i})$ are free fields of mass $m_{\left(  A\dot
{B},s\right)  ,i}~$and spin $s$ which transform according to a $\left(
2A+1\right)  (2\dot{B}+1)$ dimensional irreducible representation of the
two-fold covering $\widetilde{O(1,3)}$ of the Lorentz group which are
characterized by the SL(2,C)-"spin" $A$ and its conjugate $\dot{B}%
$\footnote{These are the famous undotted-dotted spinorial representations in
van der Waerden's notation.}. For a given spin $s$ there exists an infinite
set of pairs $(A,\dot{B}).$ The only restriction which characterizes
admissible triples $(A\dot{B},s)$ is $\left\vert A-\dot{B}\right\vert \leq
s\leq A+\dot{B}$ for $m_{i}>0$ and $A=\dot{B}$ for $m_{i}=0$.

\item For each mass and admissible triple $(A\dot{B},s)$ and each mass
$m_{\left(  A\dot{B},s\right)  ,i}$ there exist one intertwiner namely
$u^{\left(  A\dot{B},s\right)  }(s;p,s_{3})$ i.e. a rectangular matrix of
width 2s+1 and height $\left(  2A+1\right)  (2\dot{B}+1)$ (with suppressed
column indices) which convert the unitary $p$-dependent $\left(  2s+1\right)
\times\left(  2s+1\right)  ~$Wigner representation matrix (which appears in
Wigner's unitary transformation law of the irreducible $(m_{i},s)$
representation of the little group) into the $\left(  2A+1\right)  (2\dot
{B}+1)\times\left(  2A+1\right)  (2\dot{B}+1)$ matrix of the spinorial
$(A,\dot{B})$ representation of $\widetilde{O(1,3)}.$ One Wigner
representation is associated with infinitely admissible covariant
representations. The $c.c$. denotes the contribution from the antiparticle
creation contribution whose $v$-intertwiner converts the unitary equivalent
conjugate Wigner representation matrix into the $(A,\dot{B})$ Lorentz group.
In all formulas the $p~$is to be taken on the appropriate $m_{\left(  A\dot
{B},s\right)  ,i}$ mass shell.

\item The two-point function of two spinorial field is only nonvanishing if
they are mutually charge conjugate (with selfconjugate Bosons as a special
case) and hence the c-number graded commutator has the form in the third line
with $\mathbf{\Delta}^{(A\dot{B},s)}(x-y;m_{\left(  A\dot{B},s\right)  ,i},s)$
being a matrix-valued covariant polynomial acting on the scalar two-point
commutator function $\Delta(x-y,m_{\left(  A\dot{B},s\right)  ,i}).$ Since the
all $u$ and $v$ intertwiners for all admissible triples $(A\dot{B},s)$ have
been computed \cite{Wei}, one also knows all covariant graded commutator
functions. The latter can also be computed directly, thus avoiding the rather
complicated interwiners \cite{Tod}.
\end{description}

The field (\ref{admissible}) is the most general covariant free field with a
discrete mass spectrum; if the mass spectrum would be continuous the field is
called a generalized free field; the shared property is that the (graded)
commutator is a c-number commutator function. Guided by the SO(4,2) spectrum
of the hydrogen atom people in the 60s tried to find a distingushed infinite
component covariant wave function, or equivalently a "natural" infinite
component free field. We will see in a moment that string theory produces a
nontrivial model solution.

The irreducible components have an analog in any spacetime dimension $n>4,$
but the analog of the undotted/dotted formalism is more involved since the
Wigner rotations now refer to the rotation group $\widetilde{O(n-1)}$ which
has more than one Casimir invariant.

Leaving the physical interpretation aside, one may ask the mathematical
question whether there exist quantum mechanical models which can be used in
the construction of such an infinite component local field with a mass/spin
tower which is fixed by the rules of quantum mechanics. The infinite component
one-particle Hilbert space of such a model must be a subspace $H_{sub}\subset
L^{2}(\mathbb{R}^{n})\otimes H_{QM}.$For simplicity we restrict to a bosonic
situation in which the auxiliary QM is generated by a system of vector-valued
bosonic operators. The Nambu-Goto model offers a solution. One starts with the
oscillators of a quantum mechanical string and defines as $H_{QM}$ the Hilbert
space generated by the oscillator Fourier components leaving out the zero mode.

There is no unitary representation of the Lorentz group in this Hilbert space
in which each oscillator transforms according to its vector index. In fact
there is no quantum mechanical space which can support a unitary covariant
representation of the Poincar\'{e} group generating.

To see the way out it suffices to remember how one handles the finite
dimensional case with say $H_{QM}=V^{(n)}$ an n-dimensional vector space. To
obtain the correct spin one unitary representation, one passes to a subspace
in which the Lorentz group representation is isomorphic to the homogenous part
of the unitary Wigner representation. The following two relations indicate
this procedure for operators in the present case
\begin{align}
U(a,\Lambda)\left\vert p;\varphi\right\rangle  &  =e^{ipa}\left\vert \Lambda
p;u(\Lambda)\varphi\right\rangle ,~\varphi\in H_{QM}\\
U(a,\Lambda)\left\vert p;\varphi\right\rangle _{H_{sub}}  &  =e^{ipa}%
\left\vert \Lambda p;u(\Lambda)\varphi\right\rangle _{H_{sub}}+nullvector
\end{align}
where $u(\Lambda)$ denotes the natural action on the multivector indices of
the quantum mechanical states. Since the natural L-invariant inner product in
the full tensor product space is indefinite one looks (as in case of finitely
many vector indices) for a subspace $H_{sub}$ on which it becomes at least
positive semidefinite i.e. the passing to the subspace the Poincar\'{e}
transformation commutes with the condition which defines this subspace up to a
vector in $H$ of vanishing norm. The more general transformation up to a
nullvector is necessary as evidenced by the Gupta-Bleuler formalism. The last
step of passing to a positive definite inner product is canonical; one
identifies equivalence classes with respect to nullvectors.

The remaining problem is to characterize such a subspace. But this is
precisely what the $u$-intertwiner accomplish.

In the case of the infinite dimensional setting of the N-G model there two
conditions which these oscillators have to obey: the string boundary
conditions and the reparametrization invariance condition. The corresponding
quantum requirements are well-known. In the present tensor product setting
they mix the momentum of the sought object with its "internal" quantum
mechanical degrees of freedom and in this way one gets to the physical states.
The two conditions provide the additional knowledge for a master-intertwiner
which intertwines between the original covariant transformation law and a
positive metric subspace $H_{sub}$ on which the representation is
semi-unitary. As mentioned the formation of a factor space $\hat{H}_{sub}$
leads to a bona fide unitary representation.

Since all unitary representations are completely reducible and a free field in
a positive energy representation is fully determined by its two-point
function, it is clear that the resulting object is an infinite components
pointlike field and not a string in physical spacetime. The string has not
disappeared, it is encoded in the mass/spin spectrum as well as in the
irreducible component $u^{(A\dot{B},s)}$ intertwiner. Those transformations in
the oscillator space $H_{QM}$ which leave the subspace invariant and do not
implement Poincar\'{e} transformation, mix the irreducible components of the
infinite component field. This corresponds to the "wiggling" of the string but
are totally unphysical since they flip the masses and spins between different
multiplets. It is not the string in spacetime which wiggles but rather tower
over one spacetime point.

There is no reason to go into explicit computations, they can be found in
\cite{Dim}; similar calculations are also known to string theorists
\cite{Mar}. The $u^{(A\dot{B},s)}$-content of the N-G intertwiner is not
contained in those papers but it can be computed if needed. One should also
add the remark that in order to show the pointlike nature of the
localization\footnote{This means that there exists a N-G wave function-valued
distribution $u(x,..)$ which generates via test function smearing a dense set
of normalizable N-G states.} of the N-G wave functions one needs the validity
of the positive energy condition. This requires the removal of a tachyonic
component. Passing to the 10 dimensional superstring, this step becomes unnessary.

The metaphoric problem starts with declaring the localization point of the
infinite component field as the c.m. point of a string. With other words in
order to account for the quantum mechanical string one invents a string in
spacetime of which only the c. m. point is "visible". It is clear that this is
not a simple slip of the pen or a coincidental multi-slip of multi-pens;
\textit{here we are confronted with a deep misunderstanding of concepts of
local quantum physics.} Localization in quantum theories based on causal
locality, the quantum counterpart of a finite propagation speed, as opposed to
the Born probability localization in QM have a completely intrinsic
localization (the modular localization of section 4), and any attempt to
impose an interpretation from the outside (as that of a string of which only
the c. m. is visible) will sooner or later lead to the creation of a surreal
parallel world.

The only explanation I have for the present situation is that the admittedly
very subtle localization concepts, which underlie local quantum physics, have
not been mentally digested; in view of the fact that popular textbooks on QFT
are limited to commented computational recipes and obviously consider any
conceptual discussion as a waste of paper, this is lamentable but hardly
surprising. The most subtle of all structural properties of QFT is
localization which finds its most intrinsic formulation in the concept of
modular localization which is intimately connected to unitary positive energy
representations of the Poincar\'{e} group.. Perhaps at this point it is a bit
clearer becomes why a section on localization (section 4) was added in a paper
which has string theory in its sight.

The next step, namely to formulate a string theoretic interaction, did not
contribute anything in the direction of a conceptual reassessment of string
theory; to the contrary, it reinforced the metaphoric tendencies. What are
those splitting and recombining tube pictures worth, if there are no material
strings in spacetime to start with? In order to view this situation in a
historical perspective, let us briefly look back at previous struggles for
conceptual clarity in fundamental physics.

A famous episode of conceptual reassessment which immediately comes to one's
mind is the ether-relativity transition. A closer look, at how this dispute at
the turn of the 19th century evolved, shows that the problem of the ether and
the Fitzgerald-Lorentz contractions was quite innocuous as compared to that of
the conceptual status of string theory. Even if Einstein had not discovered
special relativity in 1905, sooner or later somebody would have been able to
do it. The theory had a solid observational basis, and to go beyond Lorentz it
was only necessary, permitting a philosophical simplification, to apply
Ockham's razor.

Non of the preconditions for applying Ockham's razor to string theory are met.
There are no observational facts to refer to, and on the theoretical side
there is not even a consistent concept which could serve as an ether analog,
i.e. a temporary conceptual vessel free of contradiction, which could serve as
an intermediate storage for an obviously valuable observation.

But what is there to be stored from string theory? Could it be that behind the
metaphoric camouflage there is a new idea about consistently interacting
infinite component field? The most realistic expectation after 40 years of
research without any tangible physical result is that Ockham's razor will
leave nothing; more specifically it will convert what was considered to be a
setting of particle physics into a tool in mathematics. For mathematicians the
metaphoric content is no obstacle, as long as the geometrical data are
correctly processed. The sensation to have access to a miraculous gift from
physicists discharges their imagination so that the future of string theory as
an extremely efficient factory for mathematical conjectures would be secured.
Mathematicians do not have to be concerned with the subtle relation between
geometry and material localization in local quantum physics. Apart from a very
few individuals (%
$<$%
5), they do not know the intrinsic nature of modular localization nor does it
play any role in the pursuit of their problems.

In physics the consequences are much more serious. For the first time in the
history of particle theory a whole community has entered a region in which the
capacity to distinguish between the real and the surreal has been lost. On the
one hand the idea of a string in spacetime is totally metaphoric, but on the
other hand one needs precisely this picture in order to make sense out of
interactions. This is a perfect catch 22 situation.

By having taken the old N-G model to illustrate our point of the metaphoric
versus the intrinsic, we do not want to give the impression that superstring
theory has remained on such a simple-minded old-fashioned level. Modern string
theory is a complex subject and requires a large amount of mathematical
sophistication. The use of the N-G model in this essay is only for
illustrative purposes; this is pretty much the role it plays in the first
sections in books on superstring theory. But the localization problem under
discussion is not affected by these kind of extensions to more sophisticated
implementations of the string idea. The situation is a bit like the biblical
story about Adam and Eve and the original sin which is inherited despite all
cultural enrichments.

The acceptance of the metaphoric interpretation of string theory and the kind
of thinking resulting from it has spread into parts of particle theory and
taken its toll. An example is the Klein-Kaluza idea of extra spacetime
dimensions which underwent its quantum renaissance in the entourage of string
theory. Again the metaphoric trap which blinds people to recognize the deep
link between spacetime and the localization of local quantum physics took its
toll. As a result of the very nature of inner symmetries in local quantum
physics as coming from the classification of possibilities of realizing a
local net of observable algebras, there is no way in which a spacetime
dimension can "roll up" and become an inner symmetry index. Even without
knowing the theory of how inner symmetries evolve from local nets \cite{Ha},
the horse-sense of somebody who knows about the ubiquitous presence of vacuum
polarization through localization should alert him that there is something
fishy with transporting this idea into QFT without an in-depth study how to
reconcile it with structural properties of local quantum physics..

Many physicists, especially those which have been around before string theory
took the headlines, have noticed its almost surreal appearance even if not all
of them have been able to exactly express their uneasy feeling as forcefully
as Phil Anderson, Robert Laughlin or Burt Richter. But the point of departure
into the surreal is subtle, as this essay attempts to demonstrate. What
weights even more is that the metaphors have been sanctioned by several
generations of particle theorists of the highest intellectual caliber, a very
bad precondition, certainly much worse than that at the time of ether.

Research in particle physics is not taking place in an ivory tower; it remains
under the influence of the Zeitgeist. Can one think of a better analog to the
reign of a rampant post cold war capitalism, who for decades with its false
promises managed to suffocate all ambitions for a more rational and equitable
world, and the reign of a TOE build around a metaphor?

It is no longer necessary to speculate about possible connections between
metaphors in physics and their philosophical manifestations. String theorists
not only do not deny this relation, they even take pride in spreading its
surreal content. More recently several string theorists or physicists
influenced by string theory had there outcoming in several articles and books
in which they presented their string theory supported Weltanschauung.

These articles re-enforce the take on the metaphoric surreal consequences of
string theory in this essay and they also justify the uneasy feelings which
the particle physics community outside of string theory has about what is
going on in the midst of their science. There is Susskind's world of anthropic
reasoning leading to world of multiverses\cite{Suss} as a kind of last ditch
attempt to rescue the uniqueness of superstring theory in its role as a TOE.
An even more fantastic view about the physical world can be found in articles
by Tegmark\cite{Teg} in which every mathematical theorem finds its physical
incarnation in some corner of the multiverse. It is inconceivable that
articles as \cite{Sche} would have been written without prior preparation of a
metaphor friendly area in particle theory.

It would be totally incorrect to dismiss these articles as outings of
individual oddballs of the kind previous TOEs have attracted. The philosophy
contained in those article is the logical bottom line of four decades under
the reign of an apparently mathematically consistent theory whose illusionary
aspect is not the result of an affectation of the authors with science
fiction. Rather it is an unfortunate consequence of a misunderstanding of the
most central issue of local quantum physics: the autonomous meaning of
localization. A string in QM does not know anything about placement in
spacetime, only causal theories with a maximal velocity come with this notion,
and string theory belongs to this category of theories.

In this context one should not outright dismiss the suggestion that the
immense popularity of geometry, especially differential geometry in the 70s
and 80s may have contributed to the somewhat frivolous handling of the issue
of localization. Whereas in quasiclassical approximations e.g. soliton or
monopole solution this has no effect, there is a potential danger of
metaphoric trespassing if it comes to interpretations of exact solutions.

An illustration of this point is provided by the interpretation of a chiral
observable algebra on the circle in a temperature state with respect to the
conformal Hamiltonian $L_{0}.$ The thermal correlation of such a theory
fulfill a highly nontrivial duality relation in which the upper boundary
circle of KMS of the analyticity region becomes the localizing circle of the
dual temperature chiral theory. Both the original theory and its dual "live"
(are localized) on a circle To say that this theory lives on the torus is a
metaphoric statement. The torus is an analyticity domain which relates two
models on a circle. In order to convert the upper boundary of the torus into
the living space of a chiral theory, the boundary value has to be taken in a
particular way in order to obtain the expectations of products of fields.
These (modular\footnote{Here the word modular represents stands for identities
between modular forms in complex function theory.}) dual temperature relation
are the analogs of the Nelson-Symanzik duality relation for the correlation
functions of massive two-dimensional QFT on a finite interval. In these
relations the spatial variable in correlation functions is interchanged with
in the euclidean time. Whereas the proof of these relation depends on the
validity of a strong form of the Osterwalder Schrader theorems, the chiral
case is much more complicated since space and time have been combined to a
lightray and the duality becomes a selfduality.

Evidence suggests that behind this more complicated chiral temperature duality
relation there is a noncommutative euclideanization which is related to the
modular localization structures of chiral theories on a circle, but the
mathematics which converts this evidence into a solid proof is still missing
\cite{Schrader}. Also in this case it is important to distinguish between
geometric pictures and physical localization in spacetime. Obviously there is
a Riemann surface in form of a torus associated with such a thermal chiral
theory. But this torus is not the localization region of quantum matter; the
latter only exists on the circular physical boundary. String theorists tend to
view conformal theories as worldsheet embedding of the this torus.
Geometrically this is of course correct, but from the viewpoint of physical
localization in spacetime it is a disaster.

String theorists want both, on the one hand one should think in terms of
spacetime strings. On the other hand, especially if they feel hard pressed on
this point by their critical opponents they take recourse to the excuse that
what they really had in mind is a pure S-matrix theory. In any theory which
admits objects in spacetime one needs the interceding of time-dependent
scattering theory in order to get to an S-matrix; this is the only rational
way in which "on- and off-shell" can be connected. Working on this issue with
cooking recipes would offer metaphors another breeding ground. A geometric
picture which only serves as a cooking recipe for something which one calls an
S-matrix, without having even checked those general properties with which
Stueckelberg challenged Heisenberg's S-matrix, would make both Feynman and
Stueckelberg rotate in their graves.

Having to face the surreal physical world of string theory, it interesting to
remember that at the beginning of quantum mechanics Heisenberg introduced the
concept of quantum mechanical observables (from which important restriction on
measurability and a new notion of physical reality emerged) as a precaution
against metaphoric classical traps. Let us think for a moment that QM would
have been discovered in Feynman's path integral setting. In such a situation
Heisenberg's notion of observables would have been essential for preventing a
journey into a world paved with metaphors. But once the problems of
interpretation of QM became clarified, Feynman's path integral was an
important enrichment with no danger of evocating misleading metaphors.
Unfortunately an operator version similar to the one in QM exists presently in
QFT only in a fragmentary form. Hence possible misleading interpretations
extracted from the functional integral representation are not yet banned.

It may happen that string theory will disappear through the exhaustion of its
proponents or as a result of significant changes from experiments and
observations. A move away from the Zeitgeist of the post cold war cultural
dominance of global capitalism and a change of directions away from
confrontation and exploitation may also lead to a loss of interest in grand
designs as TOEs and deplete the chances of their protagonists to obtain high
social status within the particle physics community. But considering the
enormous amount of manpower and material resources which has gone into that
project during four decades there should be a detailed account of what
precisely kept a large community of highly skilled people working on on this
project. It is hard to imagine that particle theory can have a post string
future without accounting for the problem whether string theory was more than
a collection of recipes with a misleading spacetime interpretation. A closure
only as a result of difficulties to reconcile string theory with observations
without a critical theoretical appraisal about its conceptual content would be unsatisfactory.

Apart from the introduction, the presentation has been kept on the track of a
scientifically oriented critical review. But the history as well as the
present status of string theory raises many questions whose answer has to be
looked for outside particle physics in the realm of sociology of the string
community and its coupling to the to the Zeitgeist. The following last section
attempts to address such questions.

\section{TOE, or particle theory in times of crisis}

Although there is general agreement among experts that particle physics is in
the midst of a crisis, there are diverging opinions about the underlying causes.

An often heard opinion is that the ascent of metaphoric ideas and the
increasing popularity of theories of everything (TOE)\footnote{As far as I
know the first TOE came with the German name, it was the Heisenberg Weltformal
(a nonlinear spinor theory). Pauli supported it at the beginning but later
(after Feynman's criticism) turned against it. My later Brazilean collaborator
visited Munich at the end of the 50s and got so depressed about the circus
around this Weltformel that he had doubts about his decision to go into
particle physics. Fortunately that TOE remained a local event.} is the result
of stagnation of the research on the SM caused by a lack of experimental data.
If this would be the only explanation, one could expect that forthcoming
experimental data from LHC would change directions in fundamental particle
theory research away from the TOE project towards a better understanding of QFT.

However there is no guaranty that a mere change of subject will automatically
alter the conceptual framework in which research is conducted. Whatever will
be revealed by the LHC experiments, the metaphoric style of discourse which
has been supported by string theory will not disappear only because it lost
another round against nature.

Experiments can add new facts, but as long as physical principles and
conceptual clarity do not regain the importance they had in the first four
decades of quantum theory particle physics and the sociological preferences of
theoretical research do not favor new conceptual investments in the hugely
successful, but still incomplete QFT\footnote{The level of knowledge in
present publications on such central matters as the particle-field relation
\cite{infra} has fallen below what it was decades ago.}, the present situation
will continue.

It is not an coincidence that the new less physical-conceptional and more
geometric-mathematical oriented way of doing particle theory is a post SM
phenomenon. Looking at the SM discovery, one cannot help to be impressed by
the fact that a relative modest conceptual investment which, different from
the discovery of QED two decades before did not add much to the principles of
QFT, has led to a an impressive extension of our description of nature which
has withstood the test of time for four decades.

Instead of accepting this as a result of unmerited luck, particle theorists
became intellectually arrogant and forgot that the great theoretical conquests
of the past were followed by a long struggle about problems of interpretation
which often led to heated disputes between different schools of thought as the
famous Einstein-Bohr controversy. Nothing of this kind happened in the post SM
era, even though laying claim to a TOE would require a deep conceptual
analysis more than at any previous situation. The strange proximity of
detailed and occasionally sophisticated computations and crude metaphoric
interpretations has become the hallmark of string theory and related ideas.

Most of the of the ongoing foundational work is concerned with
interpretational problems of QM. This led to an imbalance with the more
fundamental QFT. Whereas the foundational discussion in QM arrived a
nit-picking issues as the "many world interpretation" and other forms of
intellectual masturbation, a conceptual penetration of the fundamental
localization concept in QFT, which rules all issues of of physical
interpretation, has not even started. It is not easy to dismiss the thesis
that with better understanding of foundational aspects of local quantum
physics, particle theory would have shown more resistance against metaphoric
arguments and temptations to succumb to a siren's call of a TOE.

One cannot realistically expect that new experiment will change the style of
research of particle theorists whose \ thinking has been formed in the shadow
of a TOE. Too many careers have been built around ideas of a final theory and
too many prominent people have supported it in order to expect any rapid
change. A person who dedicated a good part of his/her scientific life to the
pursuit of a TOE, and in this became a renown member of the large and
influential superstring community, will find it difficult to muster the
intellectual modesty it takes to leave the limelight and start a different path.

In this respect science is not different from what happens in the
eco-political realm of society; a change of direction towards more modesty
without going through a substantial crash is not very probable.

A theory whose intrinsic properties are unknown and in which concrete
calculations, metaphoric physical arguments and subtle mathematics form an
entangled mix, presents a fertile ground for ill-defined conjectures leading
to inconclusive publications. Perhaps the most impressive illustration of how
the principal message of an interesting theoretical discovery gets lost in the
conceptual labyrinth of a TOE is the fate of the AdS-CFT correspondence. It is
quite instructive to look at this still ongoing discourse in some detail.

Already in the 60s the observation that the 15-parametric conformal symmetry
which is shared between the conformal of 3+1-dimensional compactified
Minkowski spacetime and the 4+1-dimensional Anti-de-Sitter (the opposite
constant curvature as compared to the cosmologically important de Sitter
spacetime) brought a possible field theoretic relation between these theories
into the foreground; in fact already Fronsdal \cite{Fron} suspected that a
5-dim QFTs on AdS and a 4-dim. conformal QFT share more than the spacetime
symmetry groups. But an intrinsic localization concept detached from the
chosen point-like field generators, which could have helped to convert the
shared group symmetry into a relation between two \textit{different spacetime
ordering devices} for the \textit{same abstract quantum matter substrate,} was
not yet in place. Therefore a verification of the suggestion that not only the
symmetry groups, but even the local structure of QFTs in different spacetimes
(even in the extreme case that one spacetime is a boundary of the other) may
be related, remained out of reach.

For several decades the unphysical aspect of closed timelike world lines in
the AdS\footnote{Its universal covering is however globally causal.} solution
of the Einstein-Hilbert equations was used as an argument that these equations
require the imposition of additional restrictions.

The AdS spacetime began to play an important role in particle physics when the
string theory community placed it into the center of a conjecture about a
correspondence between a particular maximally supersymmetric massless
conformally covariant Yang-Mills model in d=1+3 and a supersymmetric
gravitational model on ADS. The first paper was by J. Maldacena \cite{Ma}, who
started from a particular compactification of 10-dim. superstring theory, with
5 uncompactified coordinates forming the AdS spacetime. Since the conceptual
structure as well as the mathematics of string theory is poorly understood,
the string side was tentatively identified with one of the supersymmetric
gravity models which, in spite of its being non-renormalizable, admitted a
more manageable Lagrangian formulation and is believed to have the same
particle content. On the side of CFT string theorists placed a maximally
supersymmetric gauge theory of which calculations which verify the vanishing
of the low order beta function already existed. The vanishing of the
beta-function is a \textit{necessary} prerequisite for conformal
invariance.\ As in all Yang-Mills theories, the perturbative approach for the
correlation functions is seriously impeded by severe infrared divergencies.

The more than 5.000 follow-up papers to Maldacena's work left the conceptual
and mathematical status of the conjecture essentially unchanged. But it
elevated the Maldacena conjecture to the most important result of string
theory and its claimed connection with the still elusive quantum gravity.

The conceptual situation became somewhat more palatable after Witten
\cite{Witten} and Polyakov et al. \cite{Polya} exemplified the ideas using a
d-dimensional Euclidean functional integral setting and paying particular
attention to the $\phi^{4}$ interaction for the scalar component of the quite
involved supersymmetric Lagrangian. In this way the Maldacena conjecture
became converted into Feynman like graphical rules in terms of vertices and
propagators for both the AdS bulk to bulk as well as its conformal boundary to bulk.

The model-independent \textit{structural properties of the AdS-CFT
correspondence} came out very clearly in Rehren's \cite{Rehren}
\textit{algebraic holography}. The setting of local quantum physics (LQP) is
particularly suited for questions concerning "holography" i.e. in which a
theory is assumed as given and one wants to construct its corresponding model
on a lower spacetime associated with a boundary. Using methods of local
quantum physics one can solve such problems of isomorphisms between models in
a purely structural way i.e. without being forced to explicitly construct the
models on either side of the correspondence. QFT in its present imperfect
state of development is not capable to address detailed properties of concrete
models but it is very efficient on structural properties for classes of models
as correspondences or holographic projections.

Since generating pointlike fields are coordinatizations of spacetime-indexed
operator algebras and as such (as numerical valued coordinates in geometry)
are highly nonunique and certainly not preserved under holographic changes, an
algebraic formulation which replaces fields by the net of local algebras which
they generate, is more appropriate.

One interesting property which came out in Rehren's proof was a statement
concerning the degrees of freedom on both sides of the correspondence.
Intuitively one expects that if one starts from a Lagrangian QFT on AdS side
and the holography to the lower dimensional QFT is really a correspondence
(i.e. not a holographic \textit{projection} as in case the lightfront
holography), the resulting conformal theory should \textit{not} be of the kind
"as one knows it". A similar situation should arise in the opposite situation;
this time because there are too few degrees of freedom.

This mismatch of degrees of freedom was indeed a corollary of Rehren's
correspondence theorem. It permits the following simple computational
illustration (which does not require the more demanding mathematical setting
in \cite{Rehren}). A standard pointlike free quantum field on
AdS\footnote{Here "standard" means originating from a Lagrangian or, in more
intrinsic terms, fulfilling the time-slice property of causal propagation. A
free field is standard in this sense, a generalized free field with an
increasing Kallen-Lehmann spectral function fails to have this property.}
passes under the correspondence to a conformal generalized free field with a
continuous distribution of masses whose anomalous dimension varies with the
AdS mass parameter \cite{Du-Re}. Generalized free fields with unbounded mass
distributions (in particular this conformal generalized field as well as the
N-G generalized free field of the previous section) have a number of
undesirable properties.

Since such fields already appeared in the previous section in connection with
the N-G model (section 6), some informative historical remarks about
generalized free fields may be helpful. Such fields were introduced in the
late 50s by W. Greenberg \cite{Gre}. Their main purpose at the time was to
\textit{test the physical soundness of axioms of QFT} in the sense that if a
system of axioms allowed unphysical solutions, it needed to be further
restricted. In \cite{Ha-Sc} it was shown that generalized free fields with too
many degrees of freedom (as they arise in the mentioned illustration) lead to
a breakdown of the causal shadow (also called time-slice) property \cite{C-F}
which is the QFT analog of the classical Cauchy propagation. A related
phenomenon is the occurance of problems in defining thermal states (a maximal
"Hagedorn" temperature or worse).

Rehren's structural argument was later adapted \cite{Du-Re2}\cite{ReLec} to
the more intuitive functional integral setting (sacrificing rigor in favor of
easier communicability within the particle theory community) in order to allow
a comparison with the work of Witten \cite{Witten}, with the result that the
perturbation theory in terms of vertices and propagators agree\footnote{In yet
another implementation of the correspondence called "projective holography"
Rehren considers a formulation which uses pointlike generating fields instead
of algebras \cite{ReLec}.}. The trick which did it was a functional identity
which was very specific for AdS models; it showed that fixing functional
sources on a boundary on the one hand, and forcing the field to take a
boundary value via delta function in the functional field space on the other
hand, leads to the same result. But this clear indication in favour of
\textit{one} kind of correspondence did not make any impression on the string
community. They continued in insisting that the correspondence in Rehren's
theorem is not what they had in mind; occasionally they referred to it as the
"German AdS-CFT correspondence" thus dismissing the content of the AdS-CFT
correspondence as something which depends on geography.

The question of how such a strange situation could arise in the midst of
particle physics, the beacon of rationality, has presently no definite answer.
In lack of any other hint one guess would be that it must have been the
mismatch of degrees of freedom on both sides of the correspondence which
contravened what string theorists expected which led to the conceptual
blackout. The subordination of a theorem to the metaphoric setting of a TOE,
i.e. that mathematical rigor and conceptual cohesion are only accepted as long
as they do not get into the way of a TOE, is a unique event which has not
happened before at any other time in the history of particle physics.

This development should be deeply worrisome to the particle physics community.
Never before have there been so many (far beyond 5.000) publications with
inconclusive results on what appears an interesting, but in the broader
context of particle physics also somewhat narrow subject. In fact even
nowadays, more than one decade after the gold-digger's rush to the gold mine
of the AdS-CFT correspondence started, there is still a sizable number of
papers every month, by people looking for nuggets at the same place, but
without bringing the gravity-gauge conjecture any closer to a resolution.

Since commentaries like this run the risk of being misunderstood, let me make
perfectly clear that particle physics always was a speculative subject and it
is important that it remains this way. Therefore there is no problem
whatsoever with Maldacena's paper; it is in the best tradition of particle
physics which was always a delicate blend of a highly imaginative and
innovative contribution from one author followed by a critical analysis of
others. One should however be worried about the almost complete loss of
balance in thousands of papers trying to support a conjecture at a place which
is already occupied by a theorem without its authors even being aware of the
situation they are in.

Many of the young high energy theorists have gone into string theory in good
faith, believing that they are working at an epoch-forming paradigmatic
problem because their advisers made them think this way. Joining a globalized
community dedicated to unravel the answers to the ultimate problems of matter
and the universe with the help of a TOE is simply not a good prerequisite for
starting critical reflections.

String theory is the first theory which succeeded to dominate particle theory
without observational credentials and conceptual coherence, solely on the
claim of being a TOE. The question is how this was possible in a science,
which is considered to be the bastion of rationality and experimental
verifiability, is not easy to be answered.

Human activities even in the exact sciences were never completely independent
of the Zeitgeist. In fact if there is any sociological phenomenon to which the
frantic chase for a TOE finds it analog, it is the post-cold war reign of
globalized capitalism with its "end of history" frame of mind \cite{Fuku} and
its ideological support for insatiable greed and exploration of natural
resources. It is hard to imagine any other project in physics which would fit
the post cold war millennium spirit of power and glory and its hegemonic
claims in the pursuit of a these goals better than superstring theory: shock
and awe of a TOE against the soft conceptual power of critical thinking.

Whereas the post cold war social order has, contrary to its promises of a
better life for everybody, accentuated social differences and caused avoidable
wars and deep political divisions, the three decades long reign of the project
of a TOE in particle physics has eradicated valuable knowledge about QFT and
considerably weakened chances of finding one's way out of the present crisis.

In a science, in which the discourse at the frontier of research is as
speculative as in particle theory, one needs a solid conceptual platform from
which one can start and to which one can return if, as it is often the case,
the foray into the unknown gets stuck in the blue yonder. QFT, with its step
by step way of accumulating knowledge and its extremely strong inherent
physical principles, was able to create such a platform. The present status of
the SM does almost certainly not reflect the last word on how to formulate
renormalizable interactions in the presence of spin $\geq1$ which rather
remains one of the great future challenges of QFT.

String theory in its almost 40 year history was not able to create such a
platform; in case of failure there is nothing to fall back on. Physicists who
got directly into the string theory fray without having had the chance to get
a solid background in QFT will not be able to get out of the blue yonder since
their is no conceptually secured region from which one could look for other directions.

Many physicists entered the monoculture of string theory with only having had
a brush with a string-theoretical caricature of QFT through some computational
recipes. Hence the failure of their string theory project will not necessarily
strengthen particle theory since the greatest burden of string theory its
metaphoric style of discourse and the ability of string theorists (in
Feynman's word) to "replace arguments by excuses" will be carried on for some
time to come. Even a decisive message from the forthcoming LHC experiments
will not be able to change this situation in the short run.

String theory is the first proposal which, as the result of its Planck scale
interpretation, was effectively exempt from observational requirements. It
reached this unique status of observational invulnerability qua birth as the
result of a gigantic jump in scales of more than 15 orders of magnitude
applied to a its previous setting as a phenomenological description of certain
aspects of strong interactions. In this situation of absence of direct
observability, a fundamental theoretical discussion about its conceptual basis
would have been of the highest importance and priority. Whereas at the time of
its existence as a phenomenological dual model for strong interaction there
was yet no compelling reason to do this, when it laid claims to be the first
TOE which incorporates gravity, time was ripe for such a foundational discussion.

This chance was missed. Unlike in a similar situation around the S-matrix
bootstrap during the 60s, when renown physicists \cite{Jost} criticized some
of the more outrageous claims and later results showed that the S-matrix
principles of Poincar\'{e} invariance, unitarity, and the crossing analyticity
property admitted infinitely many "factorizing model" solutions in two
dimensions, there was no critical discussion when string theory acquired a
Planck scale interpretation. So neither the change from crossing to DHS
duality, nor the string theoretic implementation of duality and its later
claim at incorporating all known interaction (including gravity) received the
necessary critical attention.

The later geometrical enrichment rendered certain aspects of string theory
attractive to mathematicians, but did not improve observational aspects nor
its conceptual setting within relativistic QT; it mainly added a mathematical
cordon of "shock and awe" which made it more difficult even for experienced
particle theorists to get to its physical conceptual roots. This explains
perhaps why recent criticism against its hegemonic pretensions was almost
never directly aimed at the conceptual basis but instead focussed attention to
sociological and philosophical implications. A common point of attention in
all critical articles is the gross disparity between theoretical pretenses and
the total absence of observational support. Indeed superstring theory and its
claim to be a TOE created for the first time a situation in which string
theorists could acquire a high social status and have easy access to funds
without delivering tangible physical results.

Our main criticism of string theory (sections 5 and 6) was directed toward its
somewhat frivolous manner in ignoring intrinsically defined quantum concepts
as localization and the propagandistic style through which it replaces them by
metaphoric interpretations. Normally the terminology in particle physics is
set by the intrinsic property of the construct and not by metaphors which may
have been helpful in the construction. The origin of the name is inextricably
related with the first model of a relativistic string, the Nambu-Goto model.
But is was demonstrated in those two sections the quantum one-string states
and their second quantization lifting are in fact point-localized generalized
free fields. The (classical) string localization is a property of the
integrand of the N-G action does not lead to quantum strings rather the
"stringy" looking spectrum is encoded into an infinite particle/spin tower
which "sits" over one point, not a bad achievement after the futile attempts
to find interesting infinite component fields one decade before. Contrary to
statements in the string literature, the localization point is not the c. m.
of a string in spacetime, but as we have seen, the only role the classical
string configuration plays on the quantum level is that its sets the relative
normalization of the infinitely many particle components in the Kallen-Lehmann
spectral function.

Unfortunately the geometrization of recipes to implement string interactions
had a two-edged effect, it helped to organize the mathematics but was
counterproductive on problems of interpretations. As we have emphasized, the
historically important parallelism between classical and quantum localization
theory is limited to points. Without this unique stroke of luck, QFT could not
have been accessed by Lagrangian quantization and the many differential
geometric applications would not have been possible. It breaks down for
stringlike\ localization. To avoid confusion on this point, it is of course
possible to smear a pointlike physical field into a string like configuration.
But such "composite strings" are not Euler-Lagrange objects.

The lack of understanding of the intrinsic meaning\footnote{What is at issue
here is Heisenberg's insitence on "intrinsicness" by introducing the notion of
quantum \textit{observables.}} of "quantum localization" was the reason why it
took such a long time to realize that the third kind of Wigner's irreducible
positive energy representations (the zero mass "infinite spin"
representations) are localized along semiinfinite spacelike strings
\cite{MSY}. These indecomposable representations are also not associated to a
Lagrangian. On the other side of the coin we know from sections 5 and 6 that
classical N-G string Lagragians describe upon quantization an infinite
component field, the mass/spin tower "sits" over one point; and the picture of
interacting strings in spacetime is a metaphor whose true meaning in terms of
the pointlike object has remained mysterious (probably only for the reason
that it has not been investigated).

The geometric setting of string theory which in the hands of Ed Witten led to
many mathematical insights, unfortunately takes one away from an intrinsic
quantum physical understanding. It permits the creation of a mathematically
consistent, but physically metaphoric world.

Even the insistence in a pure S-matrix interpretation, which is strongly
suggested by the dual model origin of \ string theory, does not help to get it
out of a conceptual "catch 22" situation. The infinite component pointlike
quantum physical nature of the string wave function remains incompatible with
the tube picture for calculating transition amplitudes even if the latter are
"on-shell" recipes. No wonder that outside the community of TOE followers,
string theory has a profound surreal appearance; even without the more
detailed critical analysis in this essay, a particle physicist who has lived
through more healthy times senses that there is something fishy.

A pure S-matrix theory cannot be the end of theoretical research since it
contains no information about vacuum polarization; the latter is one of the
most charateristic observationally confirmed properties of QFT. Instead of the
lucid picture of a theory in spacetime in which particles and the S-matrix
arise only at asymptotically large times, string theory as an S-matrix setting
would denigrate particle physics to the result of a cooking recipe and advance
the metaphorization of particle theory.

With such a big pot of conceptual hodge podge at hand, the forcible retrieval
of uniqueness by interpreting superstring theory as a theory of a multiverse,
in which each 4-dimensional universe represents one superstring solution with
differently compactified extra dimensions seems to be only a small additional
extension into the surreal..

Apart from particle physics, the idea of a TOE also brings additional problems
of a logical-philosophical nature. Something which is unique cannot be
characterized by its properties, as shown by the futility of medieval attempts
to characterize God. Characterizations of properties of objects in QT emerge
from relations (quantum correlations) between different objects or different
parts of the same object. We have a good understanding of a theory when these
relations lead to invariants which characterize what all the members of a set
have in common, whereas the their concrete values distinguish the different
members. This viewpoint is the basis of Mermin's relational interpretation of
QM \cite{Mer} as well as the presentation of QFT as arising from the relative
positioning of a finite number of "monads" in a joint Hilbert
space.\cite{interface}.

This characterization was achieved in QFT; the result is best characterized by
an allegory: \textit{something which looks, moves, smells and sounds like an
elephant is really an elephant}\textbf{;} or in QFT: an object which fulfills
a certain list of well-known properties is really a QFT and there is no place
for a metaphor which permits a different identification. A TOE would not fit
into the Alice and Bob world of exchange of information.

But when metaphors keep hanging on for decades and the saga about the little
wiggling strings and their big cosmic counterparts even enter videos
\cite{nova}, particle physics starts to have serious problems. Almost four
decades after its inception and a subsequent confusing history, and in spite
of worldwide attention and an enormous number of publications, the conceptual
status of string theory has remained as obscure as it was at the beginning.
This conceptual opaqueness and in particular the misleading information on the
intrinsic properties of localization and the lack of interest of string theory
to critically confront its own past coupled with its ability to seduce a new
generation of physicists, may well derail particle physics for some time to come.

It would be naive to believe that an essay like this or any other kind of
critique can change the course of events. Its only value may be that it
facilitates the task of future historians and philosophers of science to
understand what really happened to particle physics at the end of the
millennium. One can be sure that there will be a lot of public interest in
work dedicated to explanations about what really happened during this strange
episode in the midst of the exact sciences.

If string theory comes to an end, the reason is most probably its inability of
making predictions or/and the exhaustion of its proponents resulting from
frantic efforts to keep the theory in the headlines. It would be regrettable
if it fades away without a final resum\'{e} because having had a project
running for almost 4 decades with an enormous expenditure of mental and
material resources one at least would like to know the precise reason why it
was abandoned and what quantum physical concepts, if any, were behind its rich
metaphoric constructions.

In the previous section we have alluded to a parallelism between the post cold
war Zeitgeist of global capitalism and its claim at ideological hegemony with
the increasing receptivity of the idea of a TOE in form of string theory. As
in all previous chapters of human history, the spirit of the times finds its
vocal presenters in a few individuals who on the one hand appear to be in
command of events but at the same are impelled by them.

Let us take notice of some of the statements coming from leading figures in
string theory.

The following statement (which is attributed to Ed Witten) underlines this
point: string theory is "\textit{the gift to the 21 century which fell by luck
already into the 20}$^{th}$\textit{ century}". It is representative for
several other similar statements whose purpose is to lift the spirit in times
of theoretical doubts and absence of any help from the observations. Such
statements have a seductive effect on newcomers and play an important role in
community building.

To see the dependence on the Zeitgeist in sharper focus compare this with
important statements from a previous epoch as the following one:
"\textit{Henceforth space by itself and time by itself shall become degraded
to mere shadows and only some kind of union of them shall remain
independent}". Everybody will immediately notice that this is a famous
quotation from Hermann Minkowski at a time when special relativity had already
acquired the status of a theory. This was a illuminative aphorism in a poetic
form which condensed the most important message of relativity. The idea of
rallying support behind a totally ill-defined incomplete project would have
been incompatible with the spirit in the first half of the previous century.

The uncritical reception can be seen from an Wikipedia excerpt which again is
representative for many other similar statement, some even to be found in
scientific publications:

\textit{In the 1990s, Edward Witten and others found strong evidence that the
different superstring theories were different limits of an unknown
11-dimensional theory called M-theory. These discoveries sparked the second
superstring revolution. When Witten named M-theory, he didn't specify what the
\textquotedblright M\textquotedblright\ stood for, presumably because he
didn't feel he had the right to name a theory which he hadn't been able to
fully describe. Guessing what the \textquotedblright M\textquotedblright%
\ stands for has become a kind of game among theoretical physicists. The
\textquotedblright M\textquotedblright\ sometimes is said to stand for
Mystery, or Magic, or Mother. More serious suggestions include Matrix or
Membrane. Sheldon Glashow has noted that the \textquotedblright
M\textquotedblright\ might be an upside down \textquotedblright
W\textquotedblright, standing for Witten. Others have suggested that the
\textquotedblright M\textquotedblright\ in M-theory should stand for Missing,
Monstrous or even Murky. According to Witten himself, as quoted in the PBS
documentary based on Brian Greene's \textquotedblright The Elegant
Universe\textquotedblright, the \textquotedblright M\textquotedblright\ in
M-theory stands for \textquotedblright magic, mystery, or matrix according to
taste.\textquotedblright}

In this case the hegemonic pretense is veiled in the form of a playful name-coquetry.

In contrast, the opening mantra for most introductory talks/articles on string
theory usually starts with an apparent matter of fact statement as:

\textit{String theory is a model of fundamental physics whose building blocks
are one-dimensional extended (strings) rather than the zero-dimensional
objects (point particles).}

But what is the meaning of this statement and where is its proof ? A speaker
who makes such intoductory remarks would probably be surprised if somebody
from the audience would question this; or worse, the questioner would perhaps
be exposed to the mocking laughter of the audience.

String theorists, being not less intelligent than other particle physicists,
know about the weakness of some of their points. But they have invested too
many years in their project in order to be able to abandon it. This created a
very different situation from earlier times when the important role of a
critical balance was considered to be paramount to keep such a highly
speculative science as particle physics in a healthy state.

Erroneous research directions are not uncommon in a speculative science as
particle theory. Even some famous physicists got lost for some time in their
life in high-flying, but at the end worthless, projects. The criticism of
their colleagues and their own strong awareness that without a critical
balance particle physics would go astray brought them back to earth. It is
well known that Pauli engaged himself for more than a year in a project which
started with Heisenberg: the (now forgotten)\footnote{One useful relic of this
otherwise forgotten attempt is the two-dimensional Thirring model.}
\textit{Weltformel} in the veil of a "nonlinear spinor theory". When he went
onto a lecture tour through the US in order to spread the new idea, he was
severely criticized by Feynman to the effect that he recognized the flaw and
abandoned the project; loosing time with finding excuses was not Pauli's style.

It is certainly true that Pauli was often abrasive with his colleagues and
even wrong (however never "not even wrong") on several occasions, albeit
mostly in an interesting and for particle physics profiting way. Certainly
neither he nor his contemporaries would have used Winston Churchill's
endurance-rallying speech\footnote{"Never, never, ..........give up".} in the
defense of a questionable theory.

The strongest illustration for the loss of this corrective critical balance is
the manner in which David Gross, representing the thinking of a large part of
the string community, has kept critics at bay by stating that superstring
theory is "\textit{the only game in town}".

There is a certain grain of (perverse) truth in string theorists self-defense
in hard pressed situations at panel discussions or interviews, when they take
recourse to the argument of David Gross that, notwithstanding all criticism,
superstring theory is the only worthwhile project. Similar to the words of a
character in a short story by the late Kurt Vonnegut's (which Peter Woit
\cite{Wo} used in a similar context):

\textit{A guy with the gambling sickness loses his shirt every night in a}

\textit{poker game. Somebody tells him that the game is crooked, rigged}

\textit{to send him to the poorhouse. And he says, haggardly, I know, I}

\textit{know. But its the only game in town.}

(Kurt Vonnegut, The Only Game in Town \textit{\cite{Vo})}

the situation in string theory is at least partially self-inflicted, although
its defenders make it appear as the result of an inevitable development in
particle physics. Self-fulfilling prophesies are not uncommon in the political
realm; a long-term derailment of particle physics as a result of the
uncritical pursuit of a TOE is however unprecedented.

For more than three decades considerable intellectual and material resources
in the form of funding research laboratories and university institutions have
been going into the advancement of string theory and this has led to a
marginalization of other promising areas. So to extend that the "no other game
in town" is an assessment of the present sociological state of particle
physics, it describes, probably unintentionally, a factually true but
unhealthy situation in particle theory.

At no time before has a proposal, which for more than four decades did not
contribute any conceptual enrichment or observable prediction to particle
physics, received that much uncritical and propagandistic support by a
worldwide. As a result superstrings and extra dimensions became part of the
popular culture \cite{nova}\cite{Green}. Their increasing importance for the
entertainment industry contrasts their ill-defined scientific status.

A successful seduction does not only require a skilled seducer, but also a
sufficient number of people who, despite all their knowledge and intellectual
capacities, permit themselves to be seduced by a TOE's glamour. Individuals
can only successfully direct tendencies which have been already latently
present. The image of a TOE has always fascinated people and there were
attempts as the S-matrix bootstrap which preceded superstring theory. But they
had a lower degree of complexity and the critical stabilizing power was strong
enough to contain them and finally send them to the dustbin of history.
Perhaps this time a gigantic failure is necessary in order to recover to the
lost critical balance and return to it the same important status which it had
in the past.

A return to the project of the SM caused by new data from LHC may remind
particle physicists that a conceptual unification cannot succeed without
starting from a solid platform of experimental data and theoretical
principles. As in successful times in the past unification should come as a
conceptual gratification at the end, but not enter as a modus operandi in the
construction of a theory.

If the last section I tried to connect the existence of the superstring
community and its claim for domination of particle physics with the help of
the ideology of a TOE to the Zeitgeist of the rule of globalized capitalism
and its ideological subordination of all ares of human life to the
maximization of profit. Globalized capitalism has passed its apogee and is
presently in the midst of a terrifying downward spiral. The message proclaimed
by its ideological defenders about the end of history \cite{Fuku} were proven
incorrect by events and the promise for a beginning of a new epoch free of
wars and social conflicts and the universal reign of democracy with a happy
life for everybody is beginning to turn sour. Less than two decades after the
doom of communism, it is the victorious capitalism which is throwing the world
into its deepest crisis.

Science has been is a very important part for the presentation of the power
and glory of the social system of globalized capitalism. A theory of
everything with promise of a glorious closure of fundamental physics at the
turn of the millennium fell on extremely fertile ground. Superstring theory
does not only enjoy strong support in the US; the European Union together with
other states is spending billions of dollars on the LHC accelerator and its
five detectors which among other things are designed for the task of finding
traces for two of string theories "predictions" namely supersymmetry and extra
dimensions. TV series on string theory such as \cite{nova} would be
unthinkable without the embedding into the millennium's power and glory Zeitgeist.

It is inconceivable, that metaphoric ideas without experimental support and
with no clear conceptual position with respect to QFT could be supported by
another Zeitgeist. Nevertheless the cause for the loss of critical judgement
in a central part of what is considered to represent the most rational science
remains somewhat of a mystery and the present attempt to link it to the spirit
of time is more descriptive than explanatory.

For a few philosophers and sociologists regressive development in the social
history of mankind do not come unexpected. Especially those of the
\textit{Frankfurt school of critical theory} anticipated dialectic changes
from enlightenment into irrationality. According to a dictum of
Horkheimer\footnote{In Horkheimer's words: \textquotedblleft!f by
enlightenment and intellectual progress we mean the freeing of man from
superstitious belief in evil forces, in demons and fairies, in blind fate --
in short, the from fear -- then denunciation of what is currently called
reason is the greatest service reason can render." cited in M. Jay, The
Dialectical Imagination. A History of the Frankfurt School and the Institute
of Social Research, 1923-1950, Univ. of California Pr., 1996, p. 253.} and
Adorno: \textit{enlightenment must convert into mythology}.

Indeed the metaphoric nature of the scientific discourse, which gained
acceptability through string theory, has presented the ideal projection screen
of mystical beliefs. No other idea coming from science had such a profound
impact on the media and on popular culture. Physics departments at renown
universities \ have become the home for a new type of scientist who spends
most of her/his time moving around spreading the message extra dimensions,
landscapes of multiverses etc. This had the effect that people outside of
science think of intergalactic journeys, star wars, UFOs, poltergeists from
extra dimensions etc. whenever they hear the word "superstring" \cite{Kaku}.

For a long time physicists were critical of suggestions that there may be a
link between the content of their science and the prevalent Zeitgeist. Indeed
the interpretation of Einstein's relativity theory in connection with the
"relativism of values" at the turn of the 19$^{th}$ century is a
misunderstanding caused by terminology; relativity is the theory of the
absolute i.e. of the observer-independent invariants.

A book by P. Foreman \cite{Fo} proposes the daring thesis that a theory in
which the \textit{classical certainty} is replace by \textit{quantum
probability} could only have been discovered in war-ridden Germany where
Spengler's book \textit{the decline of the west,} which represented the post
world war I Zeitgeist, had its strongest impact. There are reasons to be
sceptical of Foreman's arguments; I think the more palatable explanation is
that the high level of German science especially on theoretical subjects was
not at all affected by the destruction of the war and the subsequent social upheavals.

Certainly there is no \textit{direct} way in which the scientific content of
fundamental research can be influenced by processes within a society. The
probability interpretation of QT had no direct relation to the post world war
I doom and gloom inasmuch as Einstein's special relativity was not influenced
by discussion about relativism of values which had been a fashionable topic at
the beginning of the 20$^{th}$century.

The relation between science and society takes on a slightly different
perspective if one looks at the way protagonists communicated among themselves
and with the public. It is mainly in this indirect way that the Zeitgeist can
have some influence on the direction and the conceptional level of the
scientific discourse.

In particle theory the feedback is more subtle. Whereas in earlier times
leading particle physicists who were the protagonists of speculative new ideas
were also their fiercest critics, it would be difficult to imagine the
protagonists of superstrings playing this double role. Of course even things
which one cannot imagine do happen once in a while. Who would have thought a
couple of years ago that one day Alan Greenspan could come forward and
declared the post cold war kind of deregulated capitalism a failed system? Who
might be the Alan Greenspan for a failed TOE in particle physics?

\begin{acknowledgement}
I thank Fritz Coester for sharing with me his recollection about the
Stueckelberg-Heisenberg S-matrix dispute. I also recollect with pleasure
several encounters I had with a young string theorist named Oswaldo Zapata
whose growing critical attitude and disenchantment with this area of research
led him to the previous version of this article and away from string theory
into the philosophy, sociology and history of science.
\end{acknowledgement}

\section{An epilog, reminiscences about Juergen Ehlers}

My visit to Berlin at the end of May 2008 was overshadowed by the sad news
about Juergen Ehlers's sudden death. As every year, I was looking forward to
continue the interesting discussions from my previous encounter with Juergen
at the AEI in Golm.

I met Juergen for the first time at the Unversity of Hamburg in 1956 when,
initially out of curiosity while still an undergraduate, I started to frequent
Pascual Jordan%
\'{}%
s seminar on General Relativity. Juergen%
\'{}%
s philosophical background and his deep grasp of conceptual points left a
remaining impression.

This was the time when DESY was founded and the University of Hamburg became
the center of experimental particle physics in Germany. With Harry Lehmann
succeeding Wilhelm Lenz and Kurt Symanzik representing the DESY theory this
was matched on the side of particle theory. These significant events and the
fact that General Relativity at that time did not enjoy much
support\footnote{This situation only changed many decades later (at least
outside of German universities) when Juergen was able to take an active role
in the formation of a MPI-supported Relativity group in Munich which, under
his leadership, became the nucleus for the AEI in Golm.} finally changed my
mind in favor of doing my graduate work under Harry Lehmann's guidance. After
I graduated and took a position in the US I lost contact with Juergen. Several
of the relativists from the Jordan Seminar, including Juergen Ehlers and
Engelbert Schuecking had gone to the Universit of Texas in order to work with
others in a scientific program on General Relativity initiated by Alfred Schild.

Juegen's return to Germany at the beginning of the 70s and the formation of a
research group at the MPI in Munich secured the survival of General Relativity
after Pascual Jordan's retirment. The group he formed at the MPI Muenchen was
an important link between the beginnings of post war General Relativity in
Hamburg and the present AEI in Golm, which the Max Planck societey founded in
1995 as part of its extension after the German unification. Juergen Ehlers was
its founding director up to his retirement in 1998. Under his leadership the
AEI became Germanies most impressive post unification Max Planck institutes.
Since my own area of research QFT in an act of extreme shortsightedness has
been closed down in all universities of Berlin, the AEI is the only nearby
place where one can meet people with similar interests.

From that time on I met Juergen on each of my yearly visits of the FU-Berlin
from where I retired in 1998. It was easy to find topics of joint interests
because after his retirement Juergen became increasingly attracted to
foundational aspects of QM and QFT. Only after his death I learned that he
also developed quite intense personal contacts with many of the algebraic
field theorists in Germany.

Part of his interest originated certainly from his desire to understand
Jordan's role as the protagonist of QFT in more detail. When Juergen worked
under Jordan in Hamburg, quantum theory and quantum field theory were side
issues, Jordan focussed all his activities on General Relavitiy, in particular
he was looking for geophysical manifestation of a 5-dimensional extension of
Einstein's theory in which the gracitational constant became a field variable.
For this reason most participants in Jordan's seminar had little knowledge
about quantum field theory and Jordan never talked about his pathbreaking
contributions, it almost was as if he did not want to be reminded of his
glorious scientific beginnings. This was a somewhat odd situation because in
Sam Schwebers' words "Jordan is the unsung hero of QFT" whereas on the other
hand he never contributed something of comparable significance to General Relativity.

Juergen probably saw the post retirement years (as the acting director of the
AEI) as a chance to finally understand the conceptual content of QFT by
following its history. I shared the interest in the history; although I am a
quantum field theoriest, I never took the time to look at its early history
either. Juergen gave me a list of Jordan's important publication and I began
to read some. We both learned to appreciate the subtle distiction between
Jordan's and Dirac's viewpoint about relativistic Quantum Theory.

In 2004 there was an conference with international participation in Mainz
\cite{Mainz} about Jordan's contributions to the foundations of quantum
physics which was supported by the Academy of Mainz, in the organization of
which Juergen played a leading role. The talks added a lot of unknown or lost
details about the beginnings of quantum field theory.

Some time afterwards Juergen asked me about my opinion on Jordan's algebraic
construction of magnetic monopoles. I was somewhat surprised because I did not
know that Jordan had published a paper on monople quantization in the same
year as Dirac. Shortly before, I had seen a purely algebraic derivation by
Roman Jackiew. When I wrote to Roman Jackiw he was probably as surprised as I
to find his full argument with all details in Jordan's three-page paper; even
the tetrahedral drawing depicting the (in modern parlance) cohomological
aspect of the argument was there.

.Another matter of common interest was to understand the fine points in
Jordan's and Dirac's work on "transformation theory". This was the name for
the formalism by which the structural equivalence between the Heisenberg and
Schroedinger formulation of quantum mechanics was established. At the
beginning of Jordan's paper which we both red, he thanks Fritz London for
sending his results on this issue before publication and strongly praises
London's work for the clarity of his presentation. I viewed this as a
condescending remark in accordance with the social etiquette of the times, but
Juergen went to the library and really red London's article. He convinced me
that, apart from any politeness, Jordan really had profound scientific reasons
to be impressed by Fritz London's article; it is the first article which
connects the new quantum theory with the appropriate mathematical tools as the
concept of Hilbert space and "rotations" therein (unitary operators). Usually
one attributes the first connection of operators in Hilbert space with QT to
John von. Neumann. When Fritz London wrote this impressive article he was an
assistant at the Technische Hochschule in Stuttgart; unlike Jordan he was not
part of the great quantum dialog between Goettingen, Copenhagen and Hamburg.

After the Mainz conference \cite{Mainz} Juergen was engaged in what I think
was a book project about Jordan since he was compiling a selected publication
list. In this context he asked me for some advice about whether Jordan's
series of papers on what he called the\textit{ neutrino theory of light} have
the same quality as his other work and hence should also enter the selected
list. Since Jordan's contemporaries made fun of this project\footnote{The
critical reaction against the metaphoric "neutrino theory of light" title of
Jordan's papers caused his contemporaries to overlook their very interesting
bosonization/fermionization content.} (in \cite{Pais} one even finds a very
funny mocking song), Juergen had doubts about its content and was in favor of
ignoring all articles which had this title.

I looked at several of these articles and was quite surprised about their
actual content behind their metaphoric title. I finally convinced Juergen to
keep at least two of them. For somebody with a knowledge of modern concepts of
QFT these articles had nothing to do with real neutrinos or light, rather
Jordan discovered what is nowadays called the \textit{bosonization of
Fermions} (or fermionization of Bosons) which is a typical structural property
of 2-dimensional conformal field theories. Obviously Jordan saw the potential
relevance of this property, but unlike Luttinger almost 3 decades later, he
found no physical application in the context of solid state physics where the
formalism of low dimensional QFTs in certain circumstances turns out to have
useful applications. In order to "sell" his nice field theoretic result, he
used the very metaphoric "neutrino theory of light" title.

An attention-attracting title which only has a metaphoric relation with the
content would have gone well in present times, but in those days it only led
to taunts by his contemporaries. Unfortunaely the ability to distinguish
between intrinsic aspects and metaphoric presentations has suffered serious
setbacks in contemporary particle physics.

An interesting episode (in which Juergen really impressed me with his astute
critical awarenes about ongoing discussions on rather complex matters)
developed, when Juergen asked me about Maldacena's conjecture on the AdS-CFT
correspondence. He wanted to understand how there can be a conjecture about a
property which is already covered by a theorem (Rehren%
\'{}%
s theorem). I told him something similar to what I commented on this problem
in section 5 of this essay.

With some differences in the interpretation of Jordan's legacy having been
removed in previous visits, I was looking forward to return to previous
unfinished discussions on the fundamental conceptual differences between QFT
and relativistic QM which result from differences in localization and
entanglement. This time I was much better prepared than last year
\cite{interface}. I also wanted to explain the rather simple argument why
states of string theory are pointlike generated and the name "string" is a
metaphor and lacks any intrinsic meaning i.e.the arguments contained in
section 6 of this essay.

With Juergen Ehlers, the theoretical physics community in Germany looses one
of its most knowledgable and internationally renown members. It is hard to
think of anybody else who was able to combine the traditional virtues of an
analytic critical mind with a still very present curiosity about fundamental
aspects of contemporary problems.

\bigskip

\end{document}